\documentclass[english,aps,pre,twocolumn,footinbib,reprint]{revtex4}
\usepackage[latin9]{inputenc}
\usepackage{amsmath}
\usepackage{amssymb}
\usepackage{graphicx}
\usepackage{esint}
\usepackage{subfigure} 
\usepackage{color}
\usepackage{color}
\usepackage[normalem]{ulem}



\usepackage{babel}
\begin{document}

\title{
Dominance of extreme statistics in a prototype many-body Brownian ratchet
}

\author{Evan Hohlfeld}
\email{evanhohlfeld@gmail.com}
\affiliation{Lawrence Berkeley National Laboratory, Berkeley, California 94720}
\author{Phillip L. Geissler}
\email{geissler@berkeley.edu}
\affiliation{Department of Chemistry,
  University of California, Berkeley, and
  Lawrence Berkeley National Laboratory, Berkeley, California
  94720\,}
 

 \begin{abstract}
Many forms of cell motility rely on Brownian ratchet mechanisms that
involve multiple stochastic processes.  We present a computational and
theoretical study of the nonequilibrium statistical dynamics of such a
many-body ratchet, in the specific form of a growing polymer gel that pushes
a diffusing obstacle. We find that oft-neglected correlations
among constituent filaments impact steady-state kinetics and
significantly deplete the gel's density within molecular distances of its leading edge. 
These behaviors
are captured quantitatively by a self-consistent theory for extreme
fluctuations in filaments' spatial distribution.
\end{abstract}

\maketitle




Living systems have evolved many processes that exploit fluctuations
at the sub-cellular scale to transmute chemical energy into mechanical
work.  These processes, collectively referred to as Brownian ratchets,
propel cell motions such as crawling, phagocytosis, 
 and chromosome separation during
anaphase \cite{Bray}.  They generally operate by an irreversible
discrete chemical process stochastically ratcheting the advance of a
continuously diffusing degree of freedom.  For example, the
essentially irreversible polymerization of an actin filament can lock
in the diffusive advance of a load-bearing obstacle \cite{Peskin} such
as the cell membrane, a synthetic microbread \cite{Cameron,Wiesner}, or an
atomic force microscope cantilever \cite{Parekh}.

Two-body Brownian ratchets, in which rectification is driven by a
single stochastic process (e.g., a single polymerizing filament) have
been analyzed extensively. 
In particular, the basic problem of a single polymerizing filament growing against a diffusive barrier under load has been solved exactly \cite{Peskin}. 
These model problems have been widely used to discuss and rationalize the
behavior of many-body systems that are less tractable but more
directly relevant to biological motility (e.g., a collection of
polymerizing filaments that push on a diffusing obstacle) \cite{Mogilner1996,Mogilner2003,Schaus, Krawczyk,Kierfeld,Carlsson2003,Weichsel,Plastino,Sekimoto}.  
To do so,
 extant theories (and many simulations as well) have  appealed to approximations 
 that are not 
generally
justified by the underlying chemical kinetics.
 For example,  it is commonly assumed that nonequilibrium considerations are important only for the discrete, driven part of the ratcheting process; all other degrees of freedom are imagined to follow adiabatically.  In this approximation, the fluctuating obstacle is replaced by an effective, steady force acting directly on the discrete elements of the ratchet. 
This assumption of rapid equilibration 
 is explicit in some stochastic models of polymerization
ratchets \cite{Mogilner1996,Mogilner2003,Schaus, Krawczyk,Kierfeld},
implicit in some phenomenological models of actin gels
\cite{Carlsson2003,Weichsel}, and inherent to continuum models of
growing gels \cite{Plastino,Sekimoto}. 
While this effective-force approximation
can be justified on thermodynamic grounds
when external loads are sufficiently strong to stall the ratchet \cite{Kierfeld},
most biological ratchets operate far from stall conditions.

In this Letter we develop a different theory of a model $N$-filament
polymerization ratchet (see Fig. 1), one that embraces many-body
correlations in a fully nonequilibrium dynamics.  Our analysis shows
that the influence on a given filament's growth due to the ratcheting action of
its peers does not obey a simple law of large numbers.
Specifically, a mean field theory which neglects correlated fluctuations in the positions of
different filaments does not agree with exact numerical simulations.  The surprisingly
influential correlations neglected by mean field theory emerge from
the nonequilibrium nature of obstacle motion, and we find they can be
captured by a self-consistent theory for extreme
fluctuations within the polymerizing gel.  In effect, this theory
recognizes that the diffusing obstacle interacts only with the
instantaneously leading filament, an extreme member of the filament
distribution. By factorizing a two-point correlation function
involving the lead filament, we derive an 
effective equation of motion that
very successfully captures the structure and kinetics of the prototype
model.  The form of this theory, as well as its basic predictions,
should be straightforward to generalize for other many-body ratchets.

\begin{figure}
\includegraphics[width=\columnwidth]{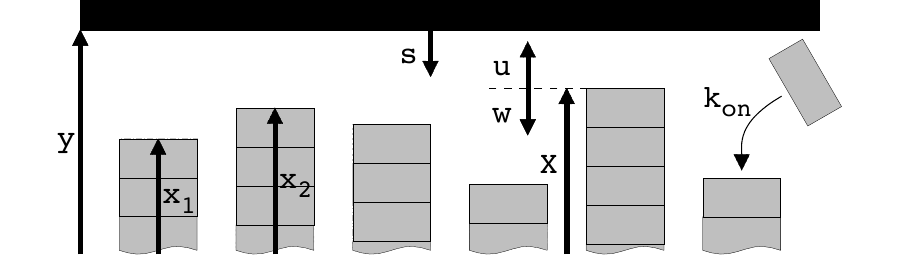}\caption{\label{fig:schematic}
  Monomers (gray rectangles) stochastically
  polymerize onto existing filaments with
  heights $x_i$, while the obstacle  (thick black line) at $y$ executes unbiased, one
  dimensional diffusion.  The coordinate $s$  is
  measured relative to $y$.  Coordinates $u$ and $w$ are measured
  relative to $X\equiv\max\left[\{x_i\}_{i=1}^N\right]$.  }
 \end{figure}

The specific model we study, sketched in Fig. \ref{fig:schematic}, is
a generalization of the $N=1$ ratchet \cite{Peskin} to the
multi-filament case.  Like that model ours focuses on the generic
physical features that are essential to its function.  In detail, we
consider $N$ parallel, straight, rigid filaments (comprising
the ``gel'') that push, via polymerization, against a diffusing obstacle with diffusion
constant $D$ and no external load. The position $x_i$ of filament tip
$i$ advances stochastically in the $+x$ direction as a conditional
Poisson process, taking discrete, irreversible steps of size $a$ with
mean rate $k_{on}$ so long as monomer addition would not penetrate the
obstacle at position $y$. The obstacle in turn diffuses freely with a
reflecting boundary condition at the leading edge of the gel, $X\equiv
\max_i x_i$. Our explicit treatment of the obstacle allows us to study
the nonequilibirum correlations inherent in the ratchet.

We imagine that the base of each filament is firmly anchored (as in the $N=1$ model),
 so that $x_i$ increases in time on a fixed
one-dimensional lattice.  Because the actin gels we have in mind are
highly disordered materials, we take the offsets among these lattices
to be randomly distributed\footnote{When the spacing between sub-lattices was held fixed as $N$ increased, qualitatively different steady-state kinetics where observed in simulation. E.g. the steady drift velocity $v$ saturates well below its kinetic limit of $k_{on}a$. }.  Adopting units of length and time such
that $a=1$ and $k_{on}=1$, the number of filaments $N$ and the 
diffusion constant $D$ (measured in units of $k_{on}a^2$)
are the only dimensionless parameters in our
model.  Biological values of this dimensionless diffusivity 
in actin-based systems range broadly,
from $D=10$ for an actin filament pushing a
patch of cell membrane \cite{Burroughs2005} to $D=10^{-2}$ for a
similar actin filament pushing the bacterium \emph{L. Monocytogenese}
through viscous cytoplasm \cite{Mogilner1996}.

We generated stochastic trajectories of our model using the continuous
time Monte Carlo (CTMC) method, which samples 
ratcheting dynamics efficiently and exactly.
Our implementation of CTMC is detailed in the \emph{Supplementary Information (SI)}.

\begin{figure}
\includegraphics[width=\columnwidth]{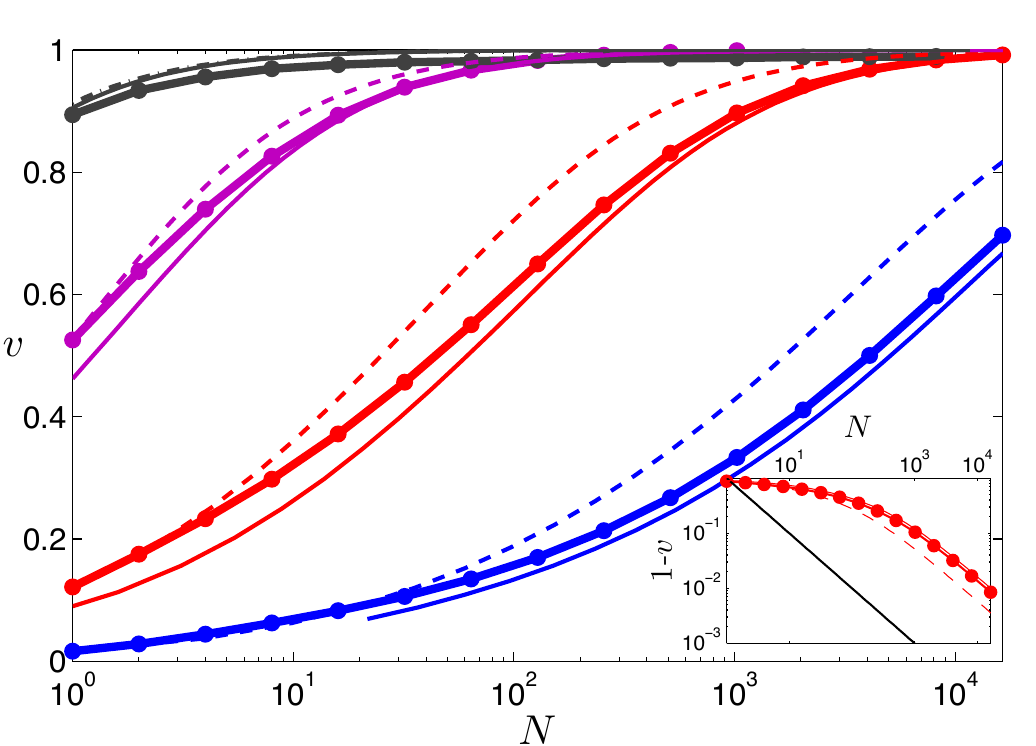}
\caption{\label{fig:v-fn}
Steady-state drift velocity $v$ as a function of the number
$N$ of pushing filaments. Results are shown
for simulations [points ($\bullet$) with thick solid lines to guide the eye] and 
predictions from MF theory (dashed lines) and from XF theory.
(solid lines).
Colors indicate different obstacle mobilities:
$D=10$
(black), $1$ (magenta), $0.1$ (red), and $0.01$ (blue). (Inset)
$1-v(N)$ (colors and markers as in the main panel) compared with $1/N$
(solid black line).}
\end{figure}

Our key numerical results include the drift velocity $v$,
 the spatially resolved average density of filament tips, and statistics of 
the lead filament's distance from the  obstacle.
The small-$N$ and large-$N$ limits of $v$
are dictated, respectively, by the 
exact solution for $N=1$ \cite{Peskin} and the average unobstructed polymerization
velocity $v=1$ as $N \rightarrow \infty$. 
Our simulation results presented in
Fig. \ref{fig:v-fn} show that the crossover between these limits is
gradual, with $v \sim \log N$ 
over a large, intermediate range of
$N$.  For $N\gg 1$ we find that $v\approx 1-\xi(D)/N,$ where $\xi(D)$
is a dimensionless function of $D$.

\begin{figure}
\includegraphics[width=\columnwidth]{
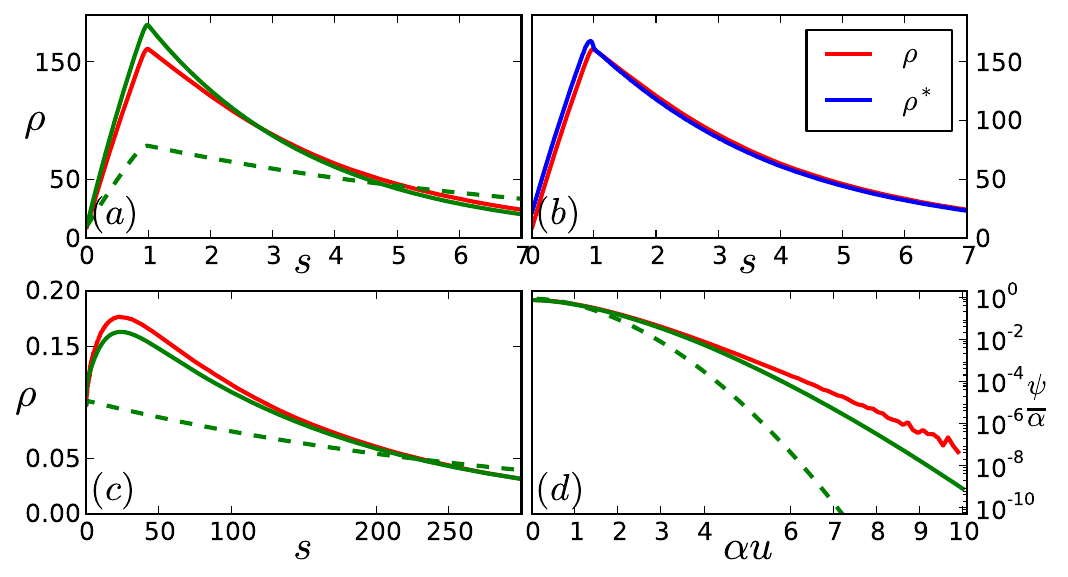}\caption{\label{fig:thry-comp}
Steady-state gel structure  as determined
from simulation and theory. In (a,b,d) $D=0.1$ and $N=600$.
(a) Filament tip density $\rho(s),$ from
  simulation (red line), from MF theory  (dashed green line), and from XF theory 
  (solid green line). (b) $\rho(s)$ (red line) and $\rho^{\star}(s)$ 
  (blue line),
both  from simulation. 
(c)  $\rho(s)$ 
for $D=10$ and $N=32$ from simulation (red), MF theory (dashed green), and XF theory (solid green). 
 (d) 
Scaled
distribution of the extreme statistic $u=y-X$, $\psi(u)$, from simulation (red), from MF theory (dashed green), and from XF theory (solid green). The scaling parameter $\alpha=[\eta(1)/D]^{1/3}\approx 10.$
}
\end{figure}

As we show in Figs. \ref{fig:thry-comp}a,d, for $N>1$ the linear density
of filament tips at distance $s$ from the obstacle, $\rho(s)$, is
marked by a molecular-scale layer of depleted filament density adjacent
to the obstacle.  In the case of low obstacle mobility, the filament density can vary by many orders of magnitude just within a single monomer distance away from the obstacle. 
Suggestively, a depleted layer of similar structure appears in the tip distribution
function of  a low-mobility $N=1$ ratchet when subjected to an external 
propulsive force.
We will
show that accurately capturing this non-monotonic density profile
requires a theory that carefully addresses extreme fluctuations
in filament density.

To clarify the relationship between steady state kinetics and
microscopic structure at the gel's leading edge,
we develop approximate analytical solutions to the master equation for the
time-dependent configurational probability $P(t,y,x_1,\dots, x_N)$ of
our $N$-filament model \footnote{Note that $P$ describes an
  \emph{ensemble} of $N$-filament ratchets with \emph{uniformly distributed} lattice
  alignments. For small $N$, the dynamics depends strongly on the relative alignment of filaments; our approach averages over these alignments.},
\begin{equation}
P_t=DP_{yy}+\sum_{i=1}^N\left[P(t,y,\dots,x_i - 1,\dots)-P\Theta (y-x_i-1)\right]
\label{eq:master}
\end{equation}
 [coordinate-name subscripts (i.e. $t$ and $y$, and later $s$ and $u$)  denote partial derivatives].
The first term on the right hand side of Eq. \eqref{eq:master} represents
free diffusion of the obstacle.  The remaining terms, which involve
the shifted coordinates $x_i\to x_i-1$ and the Heaviside function $\Theta(x)$, represent stochastic
growth of the filaments as constrained by the obstacle.  Because 
obstacle diffusion is a continuous process, the mutual impenetrability
of the obstacle and gel requires  the boundary condition $P_y(\{y=x_i\})=0$ to prevent flux of
probability to configurations that violate constraints of volume
exclusion. For $N=1$, Eq. \eqref{eq:master} is identical to the
equation of motion studied in Ref. \cite{Peskin}.

We derive several exact relationships  by evaluating  moments of
Eq. \eqref{eq:master} (detailed calculations appear in the \emph{SI}).
In particular, at steady state the average
$y$-current ($\int yP_t\,dx^N dy$) yields
the mean drift velocity $v$. Integrating
by parts
over $y$,
 we can write this moment in terms of average
structural properties, specifically the filament tip density
$\rho(s)\equiv\int_{\{x_i<y\}}\sum_i\delta(s-y+x_i)P\, dx^Ndy$ 
at a distance $s$ from the obstacle [where $\delta(x)$ is the Dirac $\delta$-function]:
\begin{equation}
v=D\rho(0).
\label{eq:velocity-formulas}
\end{equation}
According to this relation,
the effective force exerted on the obstacle, $v/D$, is
proportional to the average number of filaments \emph{in contact} with it, which is strongly shaped by multi-filament correlations.

We derive an exact equation for $\rho(s)$  from Eq. \eqref{eq:master} by multiplying both sides 
by the density operator $\sum_i\delta(s-y+x_i)$ and integrating over $x_1,\dots,x_N$,
\begin{equation}
D\rho_{ss}-D\rho_s^{(2)}(0,s)+
\rho(s+1)-\rho\Theta(s-1)=0.\label{eq:rho-eq}
\end{equation} 
The corresponding boundary condition, $\rho_s(0)=\rho^{(2)}(0,0)$, can
be obtained in similar fashion.
These results involve, but do not determine, the 
two-point correlation function 
$\rho^{(2)}(s,s') \equiv \int_{\{x_i<y\}}\sum_{i\ne j}\delta(s-y+x_i)\delta(s'-y+x_j)P\,dx^Ndy$. 

The steady state equation \eqref{eq:rho-eq} differs from a
single-filament master equation only through the term
$D\rho_s^{(2)}(0,s)$, which describes
a current of filament density induced by many-body effects.
Its form resembles the contribution $D F \rho_s(s)$
that would arise from a constant, propulsive external force $F$.  
This similarity suggests conceiving the many-filament ratchet in terms of a
single tagged filament pushing an obstacle that additionally
experiences a fluctuating force due to the remaining $N-1$ filaments.
The challenge from this perspective lies in addressing correlations
between the tagged filament's progress and fluctuations in the
effective driving force. One might naturally expect that such
fluctuations become less important with increasing $N$ and are
ultimately irrelevant in the limit $N\to \infty$. This notion
motivates a mean-field (MF) approximation to Eq.~\eqref{eq:rho-eq},
which posits a factorization of the two-point function,
$\rho^{(2)}(s,s')=\frac{N-1}{N}\rho(s)\rho(s')$, and thus neglects
correlated fluctuations in the growth of distinct filaments.  [The
coefficient $(N-1)/N$ ensures proper normalization of $\rho^{(2)}$.]
Fig. \ref{fig:thry-comp}b assesses the MF ansatz by comparing simulation results for
$\rho^\star(s)\equiv[N/(N-1)]\rho^{(2)}(0,s)/\rho(0)$ and
$\rho(s)$. These functions are indeed almost indistinguishable by eye.

The mean field factorization renders Eq.~\eqref{eq:rho-eq} simple both
to solve and to interpret. It describes a single stochastically
growing filament and an obstacle that diffuses
 under a constant  
 pulling force 
  $\mathcal{F}=\frac{N-1}{N}\rho(0)$.  The
strength of this 
force
(which represents ratcheting by the remainder of
the gel) must be determined self-consistently, through the nonlinear
boundary condition $\rho_s(0)=\mathcal{F}\rho(0)$. 
The exact solution for this effective one-dimensional system recapitulates some
of the 
qualitative 
 behaviors revealed by our simulations.

In particular, MF theory captures the emergence of a depletion layer
[i.e., large and positive density gradient $\rho_s(0)$]
for small $D$, which can be viewed as a straightforward consequence of
flux balance. When $s>1$, the tagged filament can polymerize
freely. For low obstacle mobility, the corresponding contributions to
Eq.~\eqref{eq:rho-eq} (the latter two terms on the right hand side)
nearly balance, describing steady flux of filament tip density towards
the obstacle [and consequent steady increase in $\rho(s)$ as $s$
descreases]. In $s<1$ the tagged filament stalls; the influx
$\rho(s+1)$ of polymerizing filaments in Eq.~\eqref{eq:rho-eq} must be
balanced instead by the MF drift $D\mathcal{F}\rho_s$. 
As $\rho(1)$ is large, due to the flux from $s>1$ \footnote{Using the balance $D\mathcal{F}\rho_s(0) \sim \rho(1)$ with the MF boundary condition $\rho_s(0)=\mathcal{F}\rho(0)$ and the velocity relation $v=D\rho(0) \approx D\mathcal{F}$, we compute $\rho(1) \sim D\mathcal{F}^2\rho(0) \sim D^{-2}v^3$, which is large when $D\ll v^{3/2}$.},
so must be $\rho_s(0)$.

Given the close agreement between $\rho^\star(s)$ and $\rho(s)$ in
Fig. \ref{fig:thry-comp}b, predictions of MF theory for the relationship between $N$
and $v$ are surprisingly inaccurate (see Fig. \ref{fig:v-fn}).  In particular,
the number of filaments required to sustain an average speed of
$v\approx 1/2$ errs by more than a factor of two for the lowest
obstacle mobilities we have simulated. More troublingly, this error
persists for large $N$ and
appears to grow as $D$ decreases,
i.e. as the number of  contacting filaments, $\rho(0)\sim v/D$, \emph{increases} 
[see Eq. \eqref{eq:velocity-formulas}] and
precisely where the MF approximation seems best
justified. 
Furthermore, MF theory misses 
qualitative
 features of $\rho$ when $D$ is large, most notably the persistence of the depleted layer even for $D\gg1$ (see Fig. \ref{fig:thry-comp}c). 
The inter-filament correlations neglected in MF theory,
while small in absolute magnitude, are thus highly influential for
kinetics, especially in the limit $N\to \infty$.

The failure of MF theory motivates a shift in perspective
and strategy, away from characterizing the {\em average} behavior of a
filament and towards understanding statistics of the gel's leading
edge. After all, the obstacle is 
obstructed at any moment only 
by the one filament that has grown the farthest. 
We therefore focus on the distance $u = y-X$ between the obstacle and
lead filament, whose statistical distribution $\psi(u) \equiv \int\delta(u-y+X)P\,dx^Ndy$
also directly determines steady state kinetics: $v = D \psi(0)$  (see
{\emph SI}). Our theory for the extreme fluctuations characterized by $\psi(u)$
begins with an exact but incomplete relation:
\begin{equation}
 D\psi_{uu}=\int_0^1
\left[
\Pi(u,w)\Theta(w+u-1)-\Pi(u-w+1,w)
\right]
dw,\label{eq:psi-eq-0}
\end{equation} 
together with the boundary condition $\psi_u(0)=0$.  In Eq. \eqref{eq:psi-eq-0}, the joint 
probability $\Pi(u,w) \equiv
\int\sum_{j=1}^N\delta(u-y+X)\delta(w-X+x_j)P\,dx^Ndy$ characterizes
correlations between
the position of the lead filament
and filament density fluctuations at
a lag distance $w$ behind the lead filament (see coordinate definitions in Fig.~\ref{fig:schematic}). 
Because distances in the $X$-based and $y$-based coordinate systems
are related by the equation $u+w=s$,
we can derive exact relationships between $\Pi$, $\rho$, and $\rho^{(2)}$
which clarify the relationship between the 
lead- and average-filament centered descriptions
of filament density:
\begin{subequations}
\begin{align}
\rho(s)=&\int_0^s\Pi(u,s-u)\,du\label{eq:rho-Pi}\\
\rho^{(2)}(0,s)=&\lim_{\epsilon\to0^+}\Pi(0,s+\epsilon)\label{eq:rho2-Pi}.
\end{align} 
\end{subequations}

We construct a closed set of equations through an approximate factorization 
(denoted by over bars), 
\begin{equation}
\bar{\Pi}(u,w)=\bar{\psi}(u)\bar{\sigma}(w)\label{eq:MF-2},
\end{equation}
in which the filament density
$\sigma(w)\equiv\int\sum_{j=1}^N\delta(w-X+x_j)P\,dx^Ndy$ is resolved
relative to the lead filament position. Since one filament
resides at $w=0$ by definition, $\sigma(w)$ contains a singular part
that is conveniently separated from a meaningful measure 
of the gel's internal structure,
$\eta(w)= \sigma(w) - \delta(w)$.
We will refer to the theory based on \eqref{eq:MF-2} as extreme field
(XF) theory.  The solution of XF theory for $N\gg1$ agrees
very closely with CTMC simulations,
see \emph{SI} and Figs. \ref{fig:v-fn}, \ref{fig:thry-comp}a, c, and d.  

The equations of XF theory describe fluctuations of a tagged filament
[whose distance $s$ from the obstacle is distributed according to
$\bar{\rho}(s)/N$] interacting with an
obstacle that is driven by 
another filament [notionally the lead filament, whose separation $u$ from the obstacle
  is independently distributed as $\bar{\psi}(u)$]. Since $\bar{\rho}$
and $\bar{\psi}$ are different statistics of the same population, they
are coupled by the self-consistency condition [Eqs. \eqref{eq:rho-Pi}
  and \eqref{eq:MF-2}]:
 \begin{equation}
 \bar{\rho}(s) = \bar{\psi}(s) + \int_0^s\bar{\psi}(u)\bar{\eta}(s-u)\,du.\label{eq:convolution}
 \end{equation}

In MF theory, the obstacle that impedes growth of a tagged filament is
driven by a constant force representing the rest of the gel; beyond
the steady propulsion, many-body contributions do not change the
character of this effective obstacle's motion.
Nonequilibrium dynamics of such an effective obstacle are treated
very differently in XF theory.
The distinction is most apparent in the limit that $D\gg1$ and
$N\gg1$.  Here, XF theory predicts a simple gel structure, with
filament tip density decaying exponentially behind the lead filament,
$\bar{\eta}(w) \sim D^{-1}\exp[(v-1) w]$. The corresponding sparseness
of the gel in the vicinity of the obstacle implies that $\bar{\psi}$
differs little from its $N=1$ form, just as observed in
simulations. These results for $\eta$ and $\psi$, together with the
self-consistency imposed by Eq. \eqref{eq:convolution}, yield
a zone of depleted filament density over a length scale $\sim
D$, again in close agreement with simulation. In this analysis
depletion arises in the large-$D$ limit from large excursions of the
obstacle away from the gel's leading edge, an effect that cannot be
captured by MF theory. For the case of actin and mobility $D\sim 10$,
these excursions occur on a length scale of order ten nanometers,
which in principle could be resolved experimentally using FRET
techniques.

XF and MF theories also differ in their predictions for gel structure far from the leading edge, $s\gg D$. In this region we can solve
Eqs. \eqref{eq:convolution}, \eqref{eq:rho2-Pi}, and \eqref{eq:MF-2}
for $\rho^{(2)}(0,s)$ in terms of the gradients of $\rho$,
 \begin{equation}
\bar{\rho}^{(2)}(0,s)\sim
\bar{\rho}(0)\bar{\rho}+\chi\bar{\rho}_s+\cdots,\label{eq:MF-2-r2}
\end{equation}
where $\chi\equiv\bar{\psi}(0)\int_0^\infty
u\bar{\psi}(u)\,du$. Substituting Eq. \eqref{eq:MF-2-r2} into
\eqref{eq:rho-eq} yields an equation similar to the MF equation for
$\rho$:
\begin{equation} 
D_{\rm ren}\bar{\rho}_{ss}-v\bar{\rho}_s(s)+\bar{\rho}(s+1)-\bar{\rho}\Theta(s-1)=0\label{eq:MFT},
\end{equation}
where the renormalized diffusivity $D_{ren}=(1-\chi)D$ ranges from
$\lim_{D\to 0}D_{ren}/D= 0.3156$ to $\lim_{D\to \infty}
D_{ren}=\frac{1}{2}$ (see \emph{SI}).  The similarity of these
asymptotes to 
the long-time diffusivity of the obstacle
in an $N=1$ ratchet (see \emph{SI}) suggests 
a simple physical understanding of mobility renormalization:
From the perspective of a tagged filament far from the leading edge,
the apparent random walk executed by the obstacle is not simply
characterized by the bare mobility $D$, but is instead the 
result
of independent ratcheting by the lead filament, which both induces drift
and significantly suppresses fluctuations in the obstacle's motion.

As $D\to 0$, Eq. \eqref{eq:MFT} becomes valid for all $s$ (see
\emph{SI}).  Because this result embodies the self-consistent
hypothesis of MF theory, we judge the role of extreme value statistics
for small $D$ to be less critical qualitatively than in the limit of
high mobility.
Quantitative agreement with simulations, however, is much improved
even here by the XF renormalization of $D$.  Furthermore, assuming
filament heights to be independently distributed (as suggested by the
MF ansatz) yields for small $D$ a Gaussian form for $\psi$ (see
\emph{SI}), which does not match the compressed exponential decay that
is obtained from simulations and is correctly predicted by XF theory (see
Fig. \ref{fig:thry-comp}d).

The fundamental shortcoming of MF theory for our model ratchet is the
implicit assertion that many-body growth mechanisms can be compactly
described in terms of the average behavior of individual filaments. By
contrast, XF theory recognizes that constraints imposed by the
obstacle select a sub-population of all fluctuating degrees of freedom
for special treatment (i.e. the lead filament and obstacle).
We expect that a similar focus on appropriate extreme statistics may
be helpful in more complex models where biochemical processes at the
obstacle-gel interface
(e.g., filament branching in an autocatalytic gel
\cite{Pollard2003, Wiesner})
further distinguish certain extreme
filaments.

The robust and as-yet-unobserved prediction of our 
theory and CTMC simulations 
that the filament
density drops precipitously within 
a molecular distance of the gel's leading edge
---and the many-body correlations which cause it---
certainly  
has 
significant implications for the dynamical consequences 
of 
these
processes,
 e.g. augmentation of  forces sustained by leading filaments, alteration of the transient binding between filaments and the obstacle during branching, and amplification of leading-edge fluctuations
  (which we will discuss elsewhere). 
  Continuum models of actin gels may be able to proxy the depletion affect and its consequences with modified boundary conditions.

\begin{acknowledgments}
We thank Dan Fletcher for stimulating discussions.
This work was supported by the U.S. Department of Energy, Office of Basic Energy Sciences, Division of Materials Sciences and Engineering, under Contract No. DE-ACO2-05CH11231.
\end{acknowledgments}

\bibliographystyle{plain}

\clearpage
\onecolumngrid

\appendix

\begin{center}
\Large Supplementary Information for ``Dominance of extreme statistics in a prototype many-body Brownian ratchet''

\large E. Hohlfeld and P. L. Geissler
\end{center}

\tableofcontents

\section{Outline}

In the first part of this Supplementary Information (SI), Sec. \ref{sec:CTMC}, we give more details about our numerical methods. We also present  our 
Continuous Time Monte Carlo (CTMC) 
computations for the 
obstacle's long-time
diffusivity in an $N=1$ ratchet. Sec. \ref{sec:MF-XF} of the SI presents detailed calculations based on the mean field (MF) and extreme field (XF) closures of the exact equations for the moments $\rho$ and $\psi$ of the configuration probability density $P$. Our results in this section are exact asymptotes for the forms of $\rho$ and $\psi$ (within each closure) as $N\to \infty$ and either $D\to 0$ with $N\gg D^{-2}$ or $D\to \infty$ (and $N\gg 1$). The moments themselves are formally defined in Sec. \ref{sec:moments}, wherein we also present various exact relationships between the second-order moments $\rho^{(2)}$ and $\Pi$. The exact governing equations for $\rho$ and $\psi$ are  derived in Sec. \ref{sec:exact-equations}.

\section{Continuous Time Monte Carlo (CTMC)}
\label{sec:CTMC}

\subsection{CTMC Method}

We generated stochastic trajectories of our model using the continuous
time Monte Carlo (CTMC) method, which samples 
ratcheting dynamics efficiently and exactly.
In our application of this method, time advances 
stochastically from the most recent attempted
polymerization event at $t$ to the next attempted event at
$t+\tau$. We model polymerization as a conditional Poisson process. Therefore, the random waiting time $\tau$ between polymerization
attempts is distributed as $p(\tau)=N \exp(-N\tau)$.  At each attempt,
a randomly selected filament
at position $x_i(t)$,
 polymerizes provided
it is unobstructed by the obstacle at $y(t+\tau)$,
$x_i(t)+1<y(t+\tau)$
(see Fig. \ref{fig:schematic} for coordinate definitions).  Between polymerization attempts, the obstacle
moves diffusively with a reflecting boundary condition at $X(t)\equiv\max_i x_i(t)$.
We initialized simulations by setting the obstacle location $y=a$ and 
drawing $x_i(0)$ from
a uniform distribution
in the interval $[0,a]$,
and then computed ensemble statistics by sampling
configurations at uniform time intervals following establishment of a
steady state. As $v \rightarrow 1$ in the limits of large $N$ and large $D$,
relaxation to such steady states can be very slow.

\subsection{CTMC evaluation of 
the obstacle's long-time diffusivity
for an $N=1$ ratchet}
\label{sec:Dcom}

\begin{figure}
\includegraphics[width=.75\columnwidth]{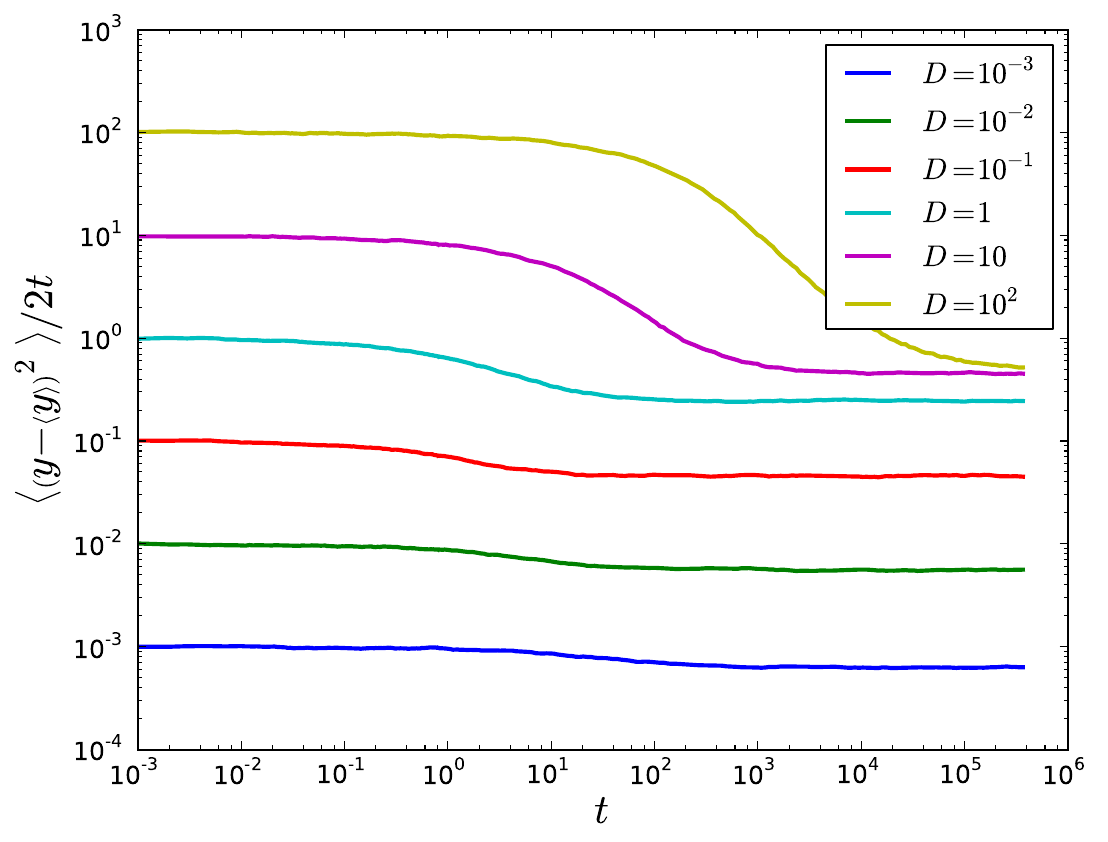}\caption{\label{fig:fig-S1}
The variance of $y$ divided by $2t$ converges to $D$ as $t\to 0$ and to $D_{com}$ as $t\to \infty$. Data are for $N=1$.
 }
 \end{figure}

\begin{figure}
\includegraphics[width=.75\columnwidth]{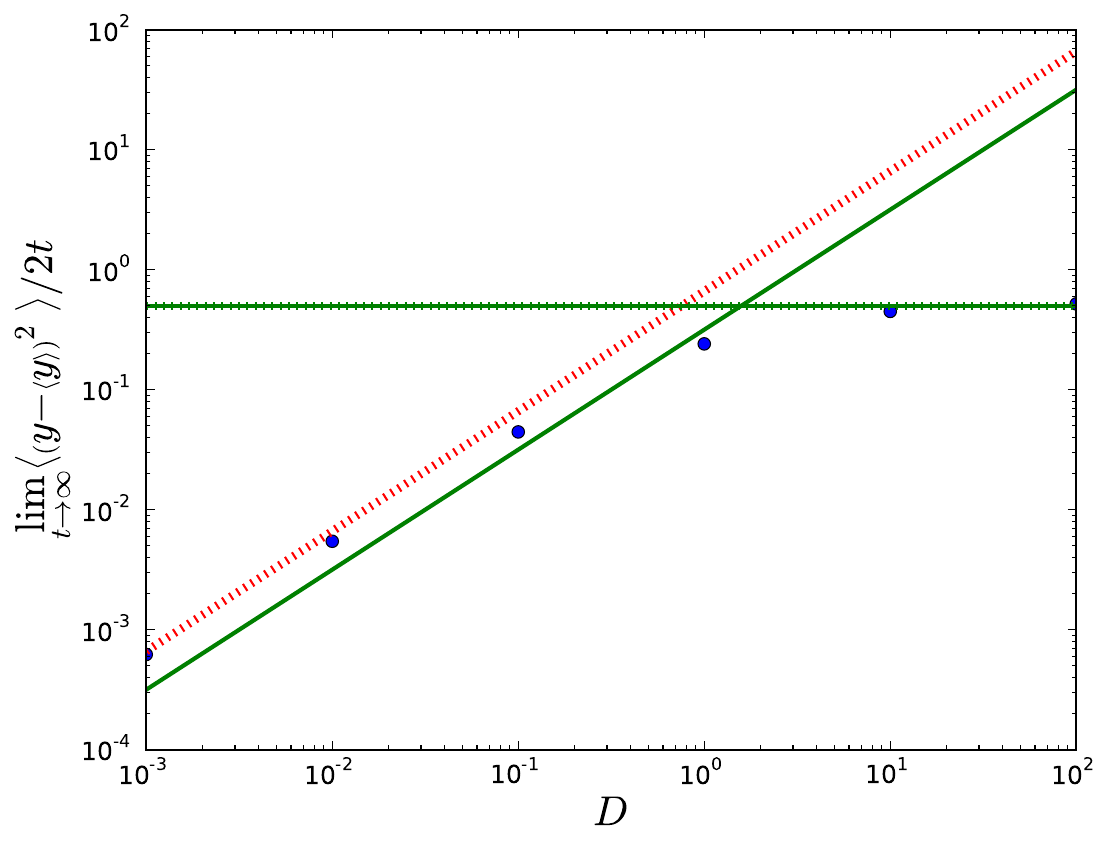}\caption{\label{fig:fig-S2}
$D_{com}$ as a function of $D$ for $N=1$ (points). Thin dotted lines are guides to the eye defined by $D_{com}=0.66\,D$ (red dotted line) and $D_{com}=\frac{1}{2}$ (green dotted line). For comparison, we also plot the  asymptotes for $D_{ren}$ as predicted by XF theory for, $D_{ren}\sim 0.3156 \, D$ as  $D\to 0$ and $D\to \frac{1}{2}$ as $D\to \infty$ (green solid lines). }
 \end{figure}

In steady-state, the relative degrees of freedom of and $N$-filament ratchet are characterized by stationary densities $\psi$ and $\rho$, etc. For an $N=1$  ratchet we have $\psi=\rho$. The center-of-mass degree of freedom, $\frac{1}{2}(x+y)$, of the ratchet never achieves a stationary distribution, and instead executes a  random walk with drift $v$ and diffusivity $D_{com}$. Whereas $v$ can be computed in various ways from the stationary distributions of the relative degrees of freedom, e.g. 
\[
v=D\psi(0)=D\rho(0)=\frac{1}{N}\int_1^\infty\rho(s)\,ds
\]
we have been unable to find any comparable formulas for  $D_{com}$. Therefore we must rely on CTMC alone to determine $D_{com}$.  Because the relative degrees of freedom are characterized by stationary, normalizable distributions at long times, $D_{com}$ can be computed from the long-time variance of any particular degree of freedom, e.g. $y$, $X$, or $x_i$. 
In particular, the value of $D_{com}$ coincides with the long-time diffusivity of the obstacle.
 Choosing $y$ and introducing the angle-bracket notation for averages
\[
D_{com}=\lim_{t\to\infty}\frac{1}{2t}\left[\iint y^2P\, dx^Ndy - \left(\iint yP\, dx^Ndy\right)^2\right]=\lim_{t\to\infty}\frac{1}{2t}\left<y-\left<y\right>\right>^2.
\]
For this particular degree of freedom, we also have
\[
D=\lim_{t\to 0}\frac{1}{2t}\left<y-\left<y\right>\right>^2.
\]
Our simulation results for $N=1$ based on 10,000 independent trials at each $D$ are detailed in Figs. \ref{fig:fig-S1} and \ref{fig:fig-S2}. 
Sampling in these simulations occurred at logarithmically spaced time intervals.


\section{Mean field (MF) and extreme field (XF) analysis\label{sec:MF-XF}}
In section \ref{sec:exact-equations} we derive that 
the filament tip density $\rho(s)$ 
solves the equation
\begin{equation}
\frac{\partial \rho}{\partial t} = D\frac{\partial^2 \rho}{\partial s^2} + D\frac{\partial }{\partial s}\rho^{(2)}(s,0)
 +\rho(s+1) - \Theta(s-1)\rho(s)\label{eq:rho_u_eq}
 \end{equation}
with the boundary condition
\begin{equation}
\frac{\partial \rho}{\partial s}(0) = \rho^{(2)}(0,0).
\end{equation} 
(See \ref{sec:moments} for a definition of $\rho$
in terms of the configurational probability $P$.) 
We also derive that 
the distribution of distance  $u=y-X$ between the lead filament and the wall, $\psi(u)$ (see Sec. \ref{sec:moments}),  
solves the equation
\begin{equation}
\frac{\partial \psi}{\partial t} = D\frac{\partial^2 \psi}{\partial u^2} + \int_0^1\Pi(u-w+1,w)\,dw - \int_{u-1}^1\Pi(u,w)\,dw\label{eq:psi_eq}
\end{equation}
with the boundary condition
\begin{equation}
\frac{\partial \psi}{\partial u}(0) = 0.
\end{equation}
Also, the instantaneous, ensemble averaged drift velocity of the obstacle is given by the exact relations
\begin{equation}
v=D\psi(0) = D \rho(0),\label{eq:v-formula}
\end{equation}
which link the values of the functions $\psi$ and $\rho$. Our objective in the next sections is to find steady-state solutions for $\rho$ and $\psi$.

\subsection{Self-consistent mean statistics}

\subsubsection{Mean-field equations}
Solving the governing equations for $\rho$ and $\psi$ [Eqs. \eqref{eq:rho_u_eq} and \eqref{eq:psi_eq}] requires knowledge of the various 
two-filament
correlation functions $\rho^{(2)}$ and $\Pi$.
(See Sec. \ref{sec:moments} for definitions of $\rho^{(2)}$ and $\Pi$ in terms of the configurational probability $P$.) 
One simple way to close Eq. \eqref{eq:rho_u_eq} is by positing the mean field like factorization
\begin{equation}
\rho^{(2)}(s,s') = \frac{N-1}{N}\rho(s)\rho(s')\label{eq:MF-factorization}.
\end{equation}
The coefficient of $(N-1)/N$ is required for proper normalization of $\rho^{(2)}$
within the factorization hypothesis.
We will refer to this factorization, which asserts the statistical independence of density fluctuations at different points in the gel, as the mean field  (MF) approximation. This factorization  seems reasonable, especially far from the obstacle $s\gg1$, as there are no direct interactions between the filaments.
Using the MF factorizations and Eq. \eqref{eq:v-formula}, we arrive at the self-consistent system
\begin{equation}
\frac{\partial \rho}{\partial t} = D\frac{\partial^2 \rho}{\partial s^2} + D\mathcal{F}\frac{\partial \rho}{\partial s}
 +\rho(s+1) - \Theta(s-1)\rho(s)\label{eq:rho_u_mft}
 \end{equation}
along the boundary condition 
\begin{equation}
\frac{\partial \rho}{\partial s}(0) =\mathcal{F}\rho(0) \label{eq:rho_bc_mft}
\end{equation}
and self-consistency condition 
\begin{equation}
\mathcal{F}=\frac{N-1}{N}\rho(0).\label{eq:mean-force}
\end{equation}
 The MF approximation transforms the complete many-filament problem encoded by Eq. \eqref{eq:rho_u_eq} into an effective problem for a single filament  in which the action of the remaining $N-1$ filaments is replaced by their mean pushing force
 $\mathcal{F}$ given by Eq. \eqref{eq:mean-force}.
  We defer solving the MF equations as their solution 
  (for any value of $D$)
  can be recovered from the XF solution when $D\ll1$ by replacing $D_{ren}\to D$.
 
 \subsubsection{Lead filament statistics in the MF approximation}
 
 The mean field approximation treats the filament tips as  independent, identically distributed random variables with distribution $\rho(s)/N$. Standard statistical theory gives a formula for the distribution of the extreme value $X=\max_i\{x_i\}$ of the multi-filament distribution function,
 \[
 P_{MF}(y-x_1,\dots,y-x_N) = \prod_{i=1}^N\frac{1}{N}\rho(y-x_i).
 \]
 Specifically, the probability that $y-X>u$ is 
 \begin{align}
 Pr(y-X>u)&= \left[\int_u^\infty \frac{1}{N}\rho(s)\,ds\right]^N\notag\\
 &= \left[1-\int_0^u \frac{1}{N}\rho(s)\,ds\right]^N.\label{eq:lead-cumulant}
 \end{align}
 The left hand side 
 of Eq. \eqref{eq:lead-cumulant}
  is one minus the cumulant function of $\psi(u)$.
 Using Euler's formula 
 for the exponential function
 to pass to the large $N$ limit, we find
 \[
 \lim_{N\to \infty}\int_u^\infty \psi(z)\,dz=e^{-\int_0^u \rho(s)\,ds}.
 \]
 Taking a derivative we have the MF expression for $\psi$,
 \begin{equation}
 \psi_{MF}(u) = \rho(u)e^{-\int_0^u \rho(s)\,ds}.\label{eq:MF-x-stats}
 \end{equation}
 
 Notice that the MF expression for $\psi$,
 Eq. \eqref{eq:MF-x-stats},
  satisfies the exact relationship $v=D\psi(0)=D\rho(0)$. Taking a derivative, we find
 \[
\frac{\partial  \psi_{MF}}{\partial u}(0) =\frac{\partial \rho}{\partial s}(0) - \rho(0)^2 =O\left[\frac{1}{N}\rho(0)\right] 
 \]
 because $\rho(s)$ satisfies the MF boundary condition, Eq. \eqref{eq:rho_bc_mft}. Hence the MF approximation to $\psi$ also satisfies the exact boundary condition for $\psi$ at $u=0$ as $N\to \infty$.
 
 We can find simple expressions for $\psi_{MF}$ when $D\to \infty$ and when $D\to 0$
 by substituting the asymptotic from of $\rho(s)$ in to Eq. \eqref{eq:MF-x-stats}. [These forms are inferred from Eq. \eqref{eq:approx-rho} below by making the substitution $D_{ren}\to D$.]
   In the first case, the mean field result for $\rho(s)\sim Ne^{-s/N}$ yields
 \[
 \psi_{MF}(u) = Ne^{-s/N}e^{-\int_0^u Ne^{-s/N}\,ds} = Ne^{-s/N}e^{e^{-s/N}-1},\quad D\gg1,
 \]
 so that $\psi$ is exponentially distributed with constant $1/N$ for large $s$. 
 That is, MF theory predicts $\psi$ is distributed according to the Gumbel Law when $D\gg1$ and $N\gg1$. This follows because $\rho(s)$ has an exponential tail.
 For comparison, XF theory also predicts that 
  $\psi(u)$ decays exponentially for large $u$,
  but
 with constant $1/D$ even as $N\to \infty$. (See Eq. \eqref{eq:psi-form} below.)
 
 In the limit $D\to 0$, $\rho(s) \sim D^{-1}+ sD^{-2}$
 for $s<1$.
On substituting this asymptotic form in Eq. \eqref{eq:MF-x-stats} we
 find that
 \[
 \psi_{MF}(u) \sim \frac{D + u}{D^2}e^{-\frac{u^2 + 2Du}{2D^2} }, \quad D\ll1,\,u\ll1.
 \]
 So $\psi_{MF}(u)$ approximately normally distributed as $D\to 0$ with variance $D^2$. We can recognize that in this limit,  $\psi$ follows  the Weibull Law with shape parameter $k=2$ for large $u$. In contrast, XF theory predicts  the compressed exponential form $\psi(u)\sim \frac{c}{\alpha}\exp[-\frac{2}{3}(\alpha u)^{3/2}]$ for $u\gg \alpha^{-1}$ (see below), which is the Weibull Law with shape parameter $k=\frac{3}{2}$ for large $u$. The 
 XF
 scale factor $\alpha\sim 1.47\,D^{-1}$ and the constant $c\approx 1.13$

 \subsection{Self-consistent extreme statistics}
 
As we discussed in the main text, the MF  approximation captures many qualitative features of growing gels, but is not quantitatively accurate, and the discrepancies between MF and exact simulations do not diminish as $N$ increases. The failure of mean field theory is that it discards information about the position of the lead filament (i.e. we do not need to solve for $\psi$
to compute $\rho$).
 In essence, the MF approximation states that the distribution of filaments at the leading edge of the gel---which are the ones actively ratcheting the obstacle---are well approximated by the extreme value statistics of uncorrelated filaments. However this approximation is qualitatively wrong.  Rather, the fluctuations in the obstacle's motion introduce correlations in the locations of the filament tips and these correlations  in turn affect the obstacle's motion. Hence, we must treat the extreme statistics of the leading edge of the gel in self consistent way.

As an alternative to MF theory, we propose closing \emph{both} the equation for $\psi$, 
Eq.\eqref{eq:psi_eq},
and the equation for $\rho$,
Eq. \eqref{eq:rho_u_eq}
 through the single factorization
\begin{equation}
\Pi(u,w)=\psi(u)\sigma(w),\label{eq:XF-factorization}
\end{equation}
which articulates
the fluctuations in the lead filament's position are uncorrected with fluctuations in the density of lagging filaments [described by the density $\sigma(w)$].
%
This closure  allows us to treat both the statistics of the  extreme filament and the mean pushing force of the lagging filaments self-consistently. The statistics of the lead filament will be consistent with the average density of filaments in the gel, while this density will be consistent with certain 
statistics of the extreme filament. 
We now discuss the overview of our theory of self-consistent extreme statistics before presenting detailed calculations in the next two subsections.

Because there is always at least one filament at the leading edge of the gel ($u=0$), the density $\sigma(u)$ has a singular part as well as an  absolutely continuous part $\eta(w)$. Formally, the Lebesgue decomposition of the measure $\sigma$ is 
\[
\sigma(w) = \delta(w) + \eta(w)
\]
where the first term on the right is a Dirac $\delta$-function and the smooth density $\eta(w)$ is normalized as
\[
\int_0^\infty \eta(w)\,dw = N-1.
\]
[See Eq. \eqref{eq:sigma-relation}  below.]
Using this definition and the XF factorization \eqref{eq:XF-factorization} in Eqs. \eqref{eq:rho_u_eq} and \eqref{eq:psi_eq} and the exact relationships between $\Pi$ and $\rho$ and $\Pi$ and $\rho^{(2)}$ given by Eqs. \eqref{eq:Pi-relation-1} and \eqref{eq:Pi-relation-2} we derive the XF equations:
\begin{subequations}
\begin{equation}
D\frac{\partial^2 {\psi}}{\partial u^2}={\psi}(u)\Theta(u-1) - {\psi}(u+1)+ 
\int_0^1\left[{\psi}(u)\Theta(w+u-1)-  
{\psi}(u-w+1)\right] {\eta}(w) dw,\label{eq:psi-eq}
\end{equation}
and 
\begin{align}
{\rho}(s)&={\psi}(s) + \int_0^\infty{\psi}(u){\eta}(s-u)\,du\label{eq:mf2a}\\
{\rho}^{(2)}(0,s)&={\psi}(0){\eta}(s)\label{eq:mf2b}
\end{align}
\label{eq:mf2}
together with Eq. \eqref{eq:rho-eq} which we reproduce here after using Eq. \eqref{eq:mf2b}, 
\begin{equation}
\frac{\partial \rho}{\partial t} = D\frac{\partial^2 \rho}{\partial s^2} - D\psi(0)\frac{\partial \eta }{\partial s}
 +\rho(s+1) - \Theta(s-1)\rho(s).\label{eq:mf2c}
\end{equation}
\end{subequations}

While the XF equations are in fact exact when $N=1$, we are interested in the $N\gg1$ asymptotic behavior of solutions to these equations, and will consider two limits, when $D\gg1$ and when $D\ll1$. The analysis of these two asymptotic regimes differ in their treatment of the convolution equation, Eq. \eqref{eq:mf2a}, in this system. In either case, for large enough $N$ and $s$ we will find that $\rho$ is slowly varying compared to $\psi$ and so Eq. \eqref{eq:mf2b} can be solved with a rapidly converging gradient expansion 
\[
\eta(s) = \rho(s) + \left[\int_0^\infty u\psi(u)\,du\right]\,\frac{\partial \rho}{\partial s} + \cdots.
\]
Substitution of this series into Eq. \eqref{eq:mf2c} shows that $\eta\sim \rho\sim C e^{-\kappa s}$ where $C$ is a normalization constant and 
\[
\kappa \sim \frac{1-v}{D_{ren}+\frac{1}{2}},\quad D_{ren} = D\left(1-\psi(0)\int_0^\infty u\psi(u)\,du\right).
\]
We  show below that $D_{ren}\to \frac{1}{2}$ as $D\to \infty$ and $D_{ren}/D\to 0.3156$ when $D\to 0$. Suggestively, these values of $D_{ren}$ are similar to the center-of-mass diffusivity of an $N=1$ ratchet, which was found from simulation to range from $\frac{1}{2}$ for $D\gg1$ to $0.66\, D$ for $D\ll1$
 (see Sec. \ref{sec:Dcom}). 
We interpret this similarity as indicating that 
the effective obstacle seen by lagging filaments consists of the lead-filament/obstacle pair.

We will see that when $D\gg1$, the exponential form of $\eta$ for $s\gg1$ is actually valid of all $s$. In this case the distribution of filament tips approaches the asymptotic form
\[
\rho(s) = \psi(s)  + \eta(0)e^{-\kappa s}\int_0^s \psi(u)e^{\kappa u}\,du,
\]
 where $\psi$ is closely approximated by its $N=1$ form. This formula for $\rho$ reveals that filament density is substantially depleted within distances $s<D$ from the obstacle. This depletion of density reflects the large excursions of the obstacle from the leading edge of the gel. Since $\psi\sim D^{-1}$ in this regime, iteration starting form these asymptotes for $\eta$ and $\rho$ converges rapidly. Whereas 
 MF
  theory predicts a monotonic form for $\rho$ in this limit, we find that the filament density at contact is roughly half its maximal value. 
  MF
theory underestimates this maximal density by roughly a factor of two.

In the other limit $D\ll1$, the average gap between the lead filament and gel scales as $D$, and thus tends to zero. In fact, we find that the distribution $\psi$ deviates substantially from the simple exponential profile it assumes when $D\gg1$. Instead, the $\psi$  is given by a scaling form that decays as an Airy function, i.e. 
\[
 \psi(\bar{u}) \sim e^{-\bar{u}^{3/2}}
\]
 (up to proportionality) for scaled distance $u[\eta(1)/D]^{1/3}=u/\alpha=\bar{u}\gg1$. As such,  the  two-term gradient expansion for $\rho^{(2)}(0,s)$  in the region $s\gg1$ remains valid for all $s$ in the limit $D\to0$ and the scaling parameter $\alpha\to0$. In this case we again find that filament density is substantially depleted close to the obstacle, but now on the scale of a monomer, i.e. unity in our dimensionless units. 
 
 The physical origin of depletion when $D\ll1$ is different from the case when $D\gg1$. When $D\ll1$, we find that  depletion results from an interaction of the the steady pushing force exerted by the large density of filaments in contact with the obstacle and the discreteness of the polymerization process. 
 A similar depleted layer appears in filament distribution for an $N=1$ ratchet with small $D$ when the obstacle experiences a pulling load. Unlike in the large $D$ asymptote, 
 MF 
 theory does capture the depletion effect in the small $D$ asymptote, but only qualitatively. 
 Because the renormalization of $D$ also modifies the boundary conditions for $\rho$ when $D\ll1$, mean field theory again underestimates the maximum filament density, 
 roughly by  a factor of  three.
 
\subsection{Asymptotic   solution of the XF equations as $N\to\infty$ and $D\to 0$}

When $s$ is large (compared to the typical distance between the lead filament at $X$ and the obstacle at $y$) and $\rho$ is slowly varying compared to $\psi$, we can solve the convolution equation,  Eq. \eqref{eq:mf2a},  for $\eta$ in terms of $\rho$ and $\psi$.
It is most convent to develop this solution by working with the  Laplace transform of ${\rho}(s)$, i.e. 
\[
\hat{\rho}(k) = \int_0^\infty e^{-ks} \rho(s)\, ds,
\]
and similarly for ${\psi}$ and for ${\eta}$. In terms of Laplace transforms, Eq. \eqref{eq:mf2a} has the form
\begin{equation}
\hat{\eta}(k) = \frac{\hat{\rho}(k)}{\hat{\psi}(k)} - 1.\label{eq:Laplace-xform}
\end{equation}
When $\psi$ is rapidly varying compared to $\rho$, $\hat{\psi}$ is slowly varying compared to $\hat{\rho}$. Supposing $\hat{\psi}$ is indeed slowly varying, we can  approximate $\hat{\psi}$ in Eq. \eqref{eq:Laplace-xform} by its low-order Taylor polynomial 
\[
\hat{\psi}(k)=\hat{\psi}(0)+k\frac{d \hat{\psi}}{d s}(0)+O(k^2),
\]
in which $\frac{d \hat{\psi}}{d s}(0)=\int_0^\infty u{\psi} (u)du.$
This approximation results in the solution for $\hat{\eta}$,
\begin{equation}
\hat{\eta}(k)=-1+\hat{\rho}(k) - k\frac{d \hat{\psi}}{d k}(0)\hat{\rho}(k) + O(k^2\hat{\rho}) \label{eq:Laplace-1}.
\end{equation}
Now inverting the Laplace transform gives
\begin{equation}
{\eta}(s) =
 {\rho}(s) + \left[\int_0^\infty u{\psi} (u)du\right]
 \frac{d {\rho}}{d s} + O\left(\frac{d^2{\rho}}{d s^2}\right),\quad s>0. \label{eq:Laplace-2}
\end{equation}
In general, the $n^{th}$ term (starting from $n=0$) in this gradient expansion for $\eta(s)$ is proportional to the $n^{th}$ moment of $\psi$ and to the $n^{th}$ derivative of $\rho$. The moments of $\psi$ can be expected to be proportional to powers of $D$, and the gradients of $\rho$ can be expected to be proportional to powers of $N^{-1}$ for large enough $N$ and $s$ (we will see that these expectations are correct). Therefore, we can expect that the series expansion for $\eta$ converges rapidly when either $N$ or $D^{-1}$ is large and $s\gg D$.

Motivated by this expected convergence, we truncate series \eqref{eq:Laplace-2} at its second term and substitute the result into Eq. \eqref{eq:mf2c} and its corresponding boundary condition, 
finding
\begin{subequations}
\begin{gather}
D_{ren}\frac{\partial^2 \rho}{\partial s^2} - v\frac{\partial \rho}{\partial s} + \rho(s+1) - \Theta(s-1)\rho(s) = 0,\\
D_{ren} \frac{\partial \rho}{\partial s}(0) = v\rho(0),
\end{gather}
\label{eq:small-D-XF}
\end{subequations}
where we have used the exact relations $\psi(0)=\rho(0)=v/D$. In these equations
\begin{equation}
D_{ren}=D\left[1-\psi(0)\int_0^\infty u\psi(u)du\right]\label{eq:D_ren}
\end{equation}
is a renormalized diffusion coefficient. 

Eqs. \eqref{eq:small-D-XF} are \emph{parametrized} by $D_{ren}$ and have the form of an effective $N=1$ ratchet under the pulling load $v/D$ where $v=D\rho(0),$ self consistently. We can solve these equations in closed form:
\begin{subequations}
\begin{equation}
\frac{1}{C}{\rho}(s) = 
\begin{cases}
e^{-\lambda s}, & s>1\\
-\frac{e^{-\lambda(s+1)}}{1-e^{-\lambda}} + \frac{1}{v\lambda}e^{-\lambda} + \left(\frac{e^{-\lambda}}{1-e^{-\lambda}} - \frac{e^{-\lambda}}{v\lambda}\right)e^{\frac{v}{D_{ren}}(s-1)},& s\le1
\end{cases}\label{eq:exact-rho}
\end{equation}
where $\lambda$ is the solution of
 the transcendental equation
\begin{equation}
D_{ren}\lambda^2 +v\lambda + e^{-\lambda} -1=0\label{eq:lambda}
\end{equation}
\end{subequations}
and where the normalization constant $C$ (i.e. the value of $N$ for given $v$ and $D_{ren}$) is fixed by self consistency. Note that this solution for $\rho$ is continuous with continuous first derivative at $s=1$.

Since $\lambda$ diminishes with increasing $N$, when $N\gg1$ we can simplify the solution for $\rho$ by expanding it in powers of $\lambda$. We find
\begin{subequations}
\begin{equation}
\frac{1}{C}{\rho}(s) =\begin{cases}
e^{-\lambda s},& s>1\\
\frac{D_{ren}}{v}\left(1-e^{\frac{v}{D_{ren}}(s-1)}\right) +  s + O(\lambda),& s>1
\end{cases}\label{eq:approx-rho}
\end{equation}
where
\begin{equation}
\lambda \sim \frac{1-v}{D_{ren} + \frac{1}{2}} + O\left[(1-v)^2D_{ren}^{-2}\right].
\end{equation}
\end{subequations}
For small $D_{ren}$ we compute the normalization constant $C$ [i.e. so that $D{\rho}(0) = v$] to be
\begin{equation}
C\sim \frac{v^2}{D D_{ren}}\label{eq:small-D-norm}
\end{equation}
and so the normalization is $N\sim C/\lambda \sim2v^2/[DD_{ren}(1-v)]$, or
\begin{equation}
v\sim 1- \frac{2}{DD_{ren} N},\quad \lambda \sim \frac{1}{DD_{ren}N}\label{eq:small-D-v-lambda}
\end{equation}
as $N\to\infty$ and $D,\,D_{ren}\to 0$.

It remains to compute $D_{ren}$ as a function of $N$ and $D$, and thus confirm that we can indeed take $N$ to infinity while holding $D_{ren}$ constant. This
calculation
 can be 
 carried out
  by analyzing Eq. \eqref{eq:psi-eq} in the limit of small $D$. In this limit, $\psi(w)$ will turn out to have no appreciable value for $w\gg D$, so we focus on the homogenous terms in Eq. \eqref{eq:psi-eq} for $w<1$ [i.e. we set $\psi(u+1)\approx0$]. Furthermore, we make the assumption that $|\frac{\partial \eta(u)}{\partial u}|\ll\eta(1)$ for $|u-1|\lesssim D$ so that we can neglect variations in $\eta$. This assumption is justified by computing from Eqs. \eqref{eq:Laplace-2} and \eqref{eq:approx-rho} that
\[
\left|\frac{1}{\eta(1)}\frac{\partial \eta}{\partial u}(1)\right|\approx \lambda\ll 1,
\]
for large enough $N$ given a value of $D_{ren}$.

With these approximations, we find that $\psi$ solves
\begin{equation}
\frac{D}{\eta(1)}\frac{\partial^2 {\psi}}{\partial u^2}= u{\psi}(u) -\int_u^\infty  
{\psi}(w) dw;\quad u\ll1,\label{eq:scaling-psi-0}
\end{equation}
with the boundary condition $\frac{\partial \psi}{\partial u}(0)=0$, and we have moved the upper limit of the integral from $u+1$ to infinity with negligible error. In this equation, the number $\eta(1)$ is suggestively placed on the the left hand side to highlight that the solution to Eq. \eqref{eq:scaling-psi-0} has the scaling form
\[
\psi(u) = \alpha \tilde{\psi}\left(\alpha u\right)
\]
where $\alpha=[\eta(1)/D]^{1/3}$.
Inspection of Eq. \eqref{eq:scaling-psi-0} shows that $\tilde{\psi}(\tilde{u})$ solves the Airy differential equation as $ \tilde{u}=\alpha u \to \infty$; hence 
\[
\tilde{\psi}(\tilde{u})\sim c\,e^{-\frac{2}{3}\tilde{u}^{3/2}},\quad \tilde{u}\gg1
\]
for some constant $c$. This form can be compared to the approximately exponential form of $\psi$ when $N=1$ or when $D\gg1$ (as we see in the next subsection).

Integrating \eqref{eq:scaling-psi-0} numerically results in the scaling form reported in the main text and yields  $c\approx 1.13\dots$. Using this 
numerically obtained
form, we computed $\chi\to 0.6844\dots$ as $D\to 0$. Hence 
\[
\lim_{D\to 0}\frac{D_{ren}}{D} =  0.3156\dots,
\]
 and on inserting this in to Eq. \eqref{eq:small-D-v-lambda} we find that $\lambda\ll1$ and thus that our small $D$ asymptote will be self consistent for $N\gg D^{-2}$. One can also check that the higher order terms in the series \eqref{eq:Laplace-2} are indeed negligible as $D\to 0$ while $N\gg D^{-2}$.

From Eqs. \eqref{eq:Laplace-2}, \eqref{eq:small-D-norm}, and \eqref{eq:small-D-v-lambda} we infer that $\eta(1)\sim C\sim (DD_{ren})^{-1}\gg1$, i.e. there is a large number of filaments stalled at $s=1$ that can contribute ratcheting as soon as the obstacle's position fluctuates a small amount. Thus  the stretched exponential profile of $\psi$ can be understood to reflect \emph{many-body} nature of ratcheting when $\sqrt{N}\gg D^{-1}$. Using 
our value for 
$\eta(1)$, we also relate the scaling parameter to $D$, $\alpha\approx 1.47\, D^{-1}$.

The solution to 
MF
%
theory can be recovered from these calculations by setting $D_{ren}=D$. We then find that MF theory underestimates the maximum filament density for a given $N$ by a factor of about three and also underestimates  $N$ for a given velocity $v$ by a factor of about two.

\subsection{Asymptotic solution of the XF equations as $N\to \infty$ and $D\to \infty$}
Now we consider the regime when $D\gg1$. Our strategy is essentially to guess  and then confirm the asymptotic solution to  Eq. \eqref{eq:mf2}. To wit, we will make several assumptions  about this solution and show that these are self-consistent a postiori. For example,  it will turn out to be consistent to assume that $\eta(s)\approx \eta(0)\sim D^{-1}$ for $0<s<1$, in which case the $\eta$-dependent term in Eq. \eqref{eq:psi-eq} is a small perturbation. Then in Eq. \eqref{eq:mf2a} we can use the $N=1$ form for $\psi$---let's call this $\psi_1$---in the convolution term, leaving a \emph{linear} system for $\eta$ and $\rho$. 
This system is:
\begin{subequations}
\begin{align}
\rho(s) &= \psi(s) + \int_0^s \psi_1(w)\eta(s-w)dw\label{eq:large-D-conv}\\
0 &= D\frac{\partial^2 \rho}{\partial s^2} - D\psi(0) \frac{\partial \eta}{\partial s} + \rho(s+1) - \Theta(s-1)\rho(s)\label{eq:large-D-rho}\\
0&=D\frac{\partial^2 \psi}{\partial u^2} + \psi(u+1) - \Theta(u-1)\psi(u) + \eta(0)\left[\int_0^1\psi(u + w)dw - \psi(u)\int_0^1\Theta(u-w)dw\right]\label{eq:large-D-psi}
\end{align}
\label{eq:large-D}
\end{subequations}
with the boundary conditions $\frac{\partial \rho}{\partial s}(0) = \eta(0)$ and $\frac{\partial \psi}{\partial u}(0) =0$. 

Let us begin 
by solving 
Eq. \eqref{eq:large-D-psi}
for $\psi$. We consider the regions $u<1$ and $u>1$ of the $u$-coordinate line separately, and then impose continuity at $s=1$ to obtain a complete asymptotic solution which is valid as $D\to \infty$.
In the region
 $u>1$, direct substitution of $\psi(u)=\exp(-q u)$ shows 
that $q$ solves the transcendental equation
\[
Dq^2 + e^{-q} + 1 + \eta(0) \frac{1-e^{-q}}{q} - \eta(0)=0.
\]
As we are interested in the case when $q\ll1$, we Taylor expand this expression and  find
\[
\left(D+\frac{\eta(0)}{6}+\frac{1}{2}\right) q^2 - \left(1+\frac{\eta(0)}{2}\right)q=O(q^3).
\]
Hence $q$ is well approximated by 
\begin{equation}
q=\frac{1+\frac{\eta(0)}{2}}{D+\frac{1}{2}+\frac{\eta(0)}{6}} + O(D^{-3}) = \frac{1}{D}\left(1+\frac{\eta(0)}{2}\right)\left(1 - \frac{1}{2D}\right) + O(D^{-3}).\label{eq:q-equation}
\end{equation}
(Notice that our assumption of small $q$ is only consistent if $\eta(0)\ll D$.) 

In
the region $u<1$,
we can approximate the slowly varying function
%
 $\psi$ by it's Taylor polynomial at $u=0$.
The boundary condition $\frac{\partial \psi}{\partial u}(0)=0$ implies the linear term in this polynomial vanishes, hence 
\[
\psi(u) = \psi(0) + \frac{1}{2}\frac{\partial^2 \psi}{\partial u^2}(0)u^2 + \frac{1}{6}\frac{\partial^3 \psi}{\partial u^3}(0) u^3 + O(u^4).
\]
Substituting this Taylor polynomial into Eq. \eqref{eq:large-D-psi} and
collecting powers of $u$, we find
\begin{subequations}
\begin{align}
D\frac{\partial^2 \psi}{\partial u^2}(0) + Ce^{-q} + \eta(0)\left[\psi(0)+\frac{1}{3}\frac{\partial \psi^2}{\partial u^2}(0) + \frac{1}{4}\frac{\partial^3\psi}{\partial u^3}(0) + \dots \right]&=0\\
\left\{D\frac{\partial^3 \psi}{\partial u^3}(0) - Cqe^{-q} + \eta(0)\left[\frac{\partial^2\psi}{\partial u^2}(0) + \dots \right] - \eta(0)\psi(0)\right\}u&=0,
\end{align}
\label{eq:s-series}
\end{subequations}
etc., where $C$ is an unknown normalization constant which will turn out to scale as $C\sim D^{-1}$.
As $\psi(0)=v/D=O(D^{-1})$ and consistency requires $\eta(0)\ll D$, we find from Eqs. \eqref{eq:s-series} that
\begin{align*}
\frac{\partial^2 \psi}{\partial u^2}(0) &= -\frac{C}{D}e^{-q} + O(\eta(0)D^{-2}),\\
\frac{\partial^3 \psi}{\partial u^3}(0) &= \frac{qC}{D}e^{-q} + O(\eta(0)D^{-2}).
\end{align*}
We obtain a formula for $\psi(0)$ by imposing that the Taylor approximation of $\psi(u)$ for $u<1$ should continuously join the exponential approximation of $\psi$ for $u>1$. This requirement results in the equation
\[
\psi(0) - \frac{C}{2D} + \frac{qC}{6D} +O(\eta(0)D^{-2} + D^{-4}) = \psi(1)  = Ce^{-q}.
\]
Up to the normalization constant $C$, we now have the solution for $\psi$:
\begin{equation}
\frac{1}{C}\psi(u) = 
\begin{cases}
e^{-qu},& u>1\\
\left(1+\frac{1}{2D} - \frac{q}{6D}\right)e^{-q} -\frac{e^{-q}}{2D}u^2  + \frac{qe^{-q}}{6D}u^3 +   O\left(\eta(0)D^{-1} + D^{-3}\right)& u\le1
\end{cases}
\label{eq:psi-form}
\end{equation}
where $q$ is defined as above in Eq. \eqref{eq:q-equation}.

To compute the normalization constant $C$ we use that $q=O(D^{-1})$ [Eq. \eqref{eq:q-equation}] to 
evaluate the integral
%
\begin{align*}
\frac{1}{C}\int_0^\infty \psi(u) du &= \frac{e^{-q}}{q} + \left(1+\frac{1}{2D} - \frac{q}{6D}\right)e^{-q} -\frac{e^{-q}}{6D} + \frac{qe^{-q}}{24D} + O\left(\eta(0)D^{-1} + D^{-3}\right)\\
&= \frac{1}{q}\left(1 +q+\frac{q}{3D} - \frac{q^2}{8D} \right)\left(1 -q + \frac{q^2}{2}\right) + O\left(\eta(0)D^{-1} + D^{-3}\right)\\
&= \frac{1}{q}\left(1 +\frac{q}{3D}  + \frac{q^2}{2}-\frac{11q^2}{24D}\right) + O\left(\eta(0)D^{-1} + D^{-3}\right),
\end{align*}
giving
\begin{equation}
C = q + O\left(\eta(0)D^{-2} + D^{-3}\right).\label{eq:psi-norm}
\end{equation}
Having the value of $C$ to this order allows us to evaluate the drift velocity $v$ to leading non-trivial order, which will be used in our self-consistency check later. We compute
\begin{align}
v &= DC\left(1 + \frac{1}{2D} - \frac{q}{6D}\right)e^{-q} + O(D^{-2} + \eta(0)D^{-1}) \notag \\
&=\left(1+\frac{\eta(0)}{2}\right)\left(1-\frac{1}{2D}\right)\left(1 + \frac{1}{2D} \right)\left(1 - \frac{1}{D} \right) + O(D^{-2} + \eta(0)D^{-1}) \notag\\
&=1 + \frac{\eta(0)}{2} - \frac{1}{D} + O(\eta(0)D^{-1}+D^{-2}).\label{eq:v-series}
\end{align}
We 
will
 also need the first moment of $\psi$ to $O(D^{-1})$. We compute
\begin{align}
\int_0^\infty u\psi(u)du &= \frac{e^{-q}}{q} + \frac{q}{2}\left(1+\frac{1}{2D}\right)\left(1-q\right) - q\frac{1-q}{8D} + O\left(\eta(0)D^{-2} + D^{-3}\right)\notag\\
&= \frac{1}{q} - 1 + q + O(\eta(0)D^{-2}+ D^{-2})\notag\\
&=D -\frac{1}{2}  + O(\eta(0) + D^{-1}).\label{eq:1st-moment}
\end{align}

Now that we have solved system Eq. \eqref{eq:large-D} for $\psi$, we next solve for $\rho$ and $\eta$. We  guess that 
\begin{equation}
\eta(s) = \eta(0) e^{-\kappa s}\label{eq:eta-form}
\end{equation}
 where $\eta(0)$ is to be determined. Using Eq. \eqref{eq:large-D-conv} 
 to express $\rho$ in terms of the known forms of $\eta$ and $\psi$ [see Eq. \eqref{eq:large-D-solution} below],
 we compute
 the following relations which are to be substituted into Eq. \eqref{eq:large-D-rho}:
\begin{subequations}
\begin{align}
\frac{\partial^2 \rho}{\partial s^2} &= \frac{\partial^2 \psi}{\partial s^2} + \eta(0)\kappa^2 e^{-\kappa s} \int_0^s \psi_1(w)e^{\kappa w}dw + \eta(0)\frac{\partial \psi_1}{\partial s}(s)  - \eta(0)\kappa \psi_1(s)\\
\rho(s) &= \psi(s) + \eta(0)e^{-\kappa s}\int_0^s \psi_1(w)e^{\kappa w}dw\\
\rho(s+1) &= \psi(s+1) + \eta(0)e^{-\kappa (s+1)} \int_0^{s+1} \psi_1(w)e^{\kappa s} dw.
\end{align}
\label{eq:large-D-forms}
\end{subequations}
We determine $\kappa$ by inserting the forms in Eqs. \eqref{eq:large-D-forms} into equation \eqref{eq:large-D-rho} and seeking a consistent solution for $s\gg1$. Recall that $\psi\sim qe^{-qs}$ for $s>1$ [see Eq. \eqref{eq:psi-form}];
therefore we find that $\kappa$ solves the equation
\[
\left(D\kappa^2 + e^{-\kappa}-1\right)\int_0^\infty\psi_1(w)e^{\kappa w}dw + \kappa D\psi(0)=0
\]
as long as $\kappa < q$ so that the integral in this equation converges.
As we are interested in the case $N\gg1$ when $\kappa\ll1$, we Taylor expand this equation in $\kappa$ and use the first moment of $\psi$ given by  Eq. \eqref{eq:1st-moment}
to compute
\[
\left[D\kappa^2 -\kappa + \frac{1}{2}\kappa^2\right]\left[1 + \kappa \int_0^\infty w\psi_1(w)dw\right] + \kappa D\psi(0)  = \kappa^2 - \left[1-D\psi(0)\right]\kappa = 0.
\]
We  hence deduce that 
\begin{equation}
\kappa = 1-v + O[(1-v)^2D^{-1}]\label{eq:kappa-series}
\end{equation}
[Recall that $D\psi(0)=v$, exactly].
 As we will see that $1-v=O(D^{-1})$, we can simplify the error estimate so that our result is $\kappa = 1-v + O(D^{-3})$.

Next we examine the small $s$ behavior of our guess. Inserting Eqs. \eqref{eq:large-D-forms} into  Eq. \eqref{eq:large-D-rho}, we find
\begin{multline}
\left\{D\frac{\partial^2 \psi}{\partial s^2} + \psi(s+1) - \Theta(s-1)\psi(s) + \eta(0)e^{-\kappa(s+1)}\int_s^{s+1}\psi_1(w)e^{\kappa w}dw - D\eta(0)\kappa \psi_1(s) +  D\psi(0)\kappa \eta(0)e^{-\kappa s}\right\} + \\
\left\{\left[D\kappa^2 +e^{-\kappa} - \Theta(s-1)\right]\eta(0)e^{-\kappa s}\int_0^s \psi_1(w)e^{\kappa s}dw + D\frac{\partial \psi_1}{\partial s}\right\}=0.
\end{multline}
Here the first group of terms tends to a finite value as $s\to 0$ and the second group tends to zero in the same limit.
  We use that $\psi_1(s) = \psi_1(0) + \frac{1}{2}\frac{\partial^2\psi}{\partial s^2}(0)s^2 + \dots$, where $\psi(0)=O(D^{-1})$ and $\frac{\partial^2 \psi}{\partial s^2}=O(D^{-2})$ [see Eq. \eqref{eq:psi-form}], as well as the relation for $\kappa$ [Eq. \eqref{eq:kappa-series}] to find that 
  \begin{multline*}
\left\{D\frac{\partial^2 \psi}{\partial s^2} + \psi(s+1) - \Theta(s-1)\psi(s) + \eta(0)e^{-\kappa}\int_0^{1}\psi_1(s+w)e^{\kappa w}dw - D\eta(0)\kappa \psi_1(s) +  D\psi(0)\kappa \eta(0)e^{-\kappa s}\right\} \\
+O(D^{-2}s + D^{-1}\eta(0)s)e^{-qs}=0.
\end{multline*}
Expanding this equation in powers of  $\kappa$, and adding and subtracting the final term in braces below [which is $O(\eta(0)D^{-1}se^{-qs})$], we compute
  \begin{multline*}
\left\{D\frac{\partial^2 \psi}{\partial s^2} + \psi(s+1) - \Theta(s-1)\psi(s) + \eta(0)\int_0^{1}\psi_1(s+w) - \psi(s)\Theta(w-s)dw\right\} \\+O( \kappa \eta(0) +  D^{-2} + D^{-1}\eta(0))se^{-qs}=0.
\end{multline*}
We recognize the term in braces as Eq. \eqref{eq:large-D-psi}, which is zero; therefore, the correction to our exponential guess for  $\eta$ (call this correction $\delta \eta$) can be found by solving an inhomogeneous first order ordinary differential equation, i.e.
\[
v\frac{\partial \delta\eta}{\partial s} = O( \kappa \eta(0) +  D^{-2} + D^{-1}\eta(0))se^{-qs}.
\] Since we will see that $\eta(0)\sim \kappa\sim D^{-1}$, the correction to $\eta$ is $O(D^{-2}e^{-qs})$. Then, by iteration with Eq. \eqref{eq:large-D-conv}, the corresponding correction to $\rho=O(D^{-3}e^{-qs})$. This correction is small compared to our leading order solution for $\rho$, which is simply given by Eq. \eqref{eq:large-D-conv} with the substitution of Eq. \eqref{eq:eta-form}:
\begin{equation}
\rho(s) = \psi(s) + \eta(0)e^{-\kappa s}\int_0^s\psi_1(w)e^{\kappa w}dw + O(D^{-3}e^{-\kappa s}).\label{eq:large-D-solution}
\end{equation}
Again, $\kappa$ was given in Eq. \eqref{eq:kappa-series}, and the error estimate is uniform in $s$. One can easily check that this solution for $\rho$ satisfies the boundary and self-consistency conditions exactly. 

Finally, we relate $\eta(0)$ to $N$ by requiring the normalization of $\eta$:
\begin{equation}
\int_0^\infty \eta(0)e^{-\kappa s}ds = \frac{\eta(0)}{\kappa} = \frac{\eta(0)}{1-v} + O(\eta(0)D^{-4}) = N-1,\label{eq:large-D-self-consistency-1}
\end{equation}
which is closed with the relation \eqref{eq:v-series} computed from our analysis of $\psi$:
\begin{equation}
1-v=  \frac{1}{D}- \frac{\eta(0)}{2} + O(\eta(0)D^{-1} + D^{-2}).\label{eq:large-D-self-consistency-2}
\end{equation}
%
%
Solving Eqs. \eqref{eq:large-D-self-consistency-1} and \eqref{eq:large-D-self-consistency-2} for $\eta(0)$ we find
\begin{equation}
\eta(0) = \frac{1-\frac{1}{N}}{1+\frac{1}{N}}\frac{2}{D} = \frac{2}{D} -\frac{4}{ND} + O(N^{-2}+D^{-2}) \label{eq:d-large-D-eta}
\end{equation}
and
\begin{equation}
v=1-\frac{2}{ND} + O(N^{-2}+ D^{-2})\label{eq:large-D-v}
\end{equation}
Inspection of the solution for $\rho$, Eq. \eqref{eq:large-D-solution}, shows that as $\kappa\to 0$, the maximum value of $\rho$ approaches $\eta(0)$. This value is achieved at a distance $s\sim D$ from the obstacle, whereas the value at contact is $\rho(0)=v/D$. We see that as $N\to \infty$, $\eta(0)/\rho(0)\to2$, hence extrapolation from the  maximum value of $\rho$ overestimates the filament density within molecular distances of the obstacle by a factor of about two.

Note that our original assumptions that $\kappa<q\ll1$ and $|\frac{\partial \eta}{\partial s}|\ll\eta(s)$ are automatically satisfied for any $N>1$ as long as $D\gg1$.

\section{Moments of $P$\label{sec:moments}}
\subsection{One-filament densities}
The statistical properties of an ensemble of $N$ growing filaments are characterized by the probability density 
\[
P\left(x_1,\dots, x_N ,y,t\right).
\]
The coordinates $x_i$ are the positions of the filament tips, $y$ is the position of the obstacle, and $t$ is time. In our theoretical approach, we focus on only a few  moments of $P$. To define these moments, we first introduce the location of the leading edge of the gel as
\[
X=\max\left[\{x_i\}_{i=1}^N\right].
\]
 We will analyze the density  of filament tips at a distance $s$ from the obstacle at $y$
\begin{equation}
\rho(s) = \iint \sum_{i=1}^N\delta(s-y+x_i)P\Theta(y-X)\,dx^Ndy\label{eq:rho-def}
\end{equation}
 the distribution of distance $u$ between the obstacle and lead filament at $X$,
\begin{equation}
\psi(u) = \iint\delta(u-y+X)P\Theta(y-X)\,dx^Ndy,\label{eq:psi-def}
\end{equation}
and the density of  filament tips at a distance $w$ from the leading edge of the gel at $X$,
\begin{equation}
\sigma(w) = \iint\sum_{j=1}^N\delta(w-X+x_j)P\Theta(y-X)\,dx^Ndy.\label{eq:sigma-def}
\end{equation}
Note that each of these time dependent densities is zero for negative values of its argument.

Because there is always one filament present at $X$, the density $\sigma$ has a singular part at $w=0$. We can isolate this singular part from the absolutely continuous part of $\sigma$  by separating the domain of integration into the sets $\{X=x_k\}_{k=1}^N$, i.e. into the $N$ sectors of $\mathbb{R}^N$ where filament each filament $i$ is the lead filament. Noting that the sets with $\{X=x_k=x_i,\,i\ne j\}$ have zero measure, we compute
\begin{align}
\sigma(w,t) =& \sum_{k=1}^N\iint\sum_{j=1}^N\delta(w-x_k+x_j)P\Theta(y-x_k)\chi_{\{X=x_k\}}\,dx^Ndy\notag\\
&=\sum_{k=1}^N\iint\delta(w)P\Theta(y-x_k)\chi_{\{X=x_k\}}\,dx^Ndy +\sum_{k=1}^N\iint\sum_{j\ne k}\delta(w-x_k+x_j)P\Theta(y-x_k)\chi_{\{X=x_k\}}\,dx^Ndy\notag\\
&= \delta(w) + \eta(w,t).\label{eq:sigma-relation}
\end{align}
Here we have used the characteristic function of a set $A$, which is defined as
\[
\chi_{A}(x)=
\begin{cases}
1 & x\in A\\
0& x\notin A.
\end{cases}
\]
We have also defined the density of lagging filaments
\[
\eta(w,t) = \sum_{k=1}^N\iint\sum_{j\ne k}\delta(w-x_k+x_j)P\Theta(y-x_k)\chi_{\{X=x_k\}}\,dx^Ndy,
\]
which satisfies the normalization $\int_0^\infty \eta(w)\,dw = N-1$.

\subsection{Two-filament densities}

The time evolution equations for the moments introduced above  involve the two-filament density functions,
\begin{equation}
\rho^{(2)}(s,s')=\iint \sum_{i=1}^N\sum_{j\ne i}\delta(s-y+x_i)\delta(s'-y+x_j)P\Theta(y-X)\,dx^Ndy,\label{eq:rho2-def}
\end{equation} 
and
\begin{equation}
\Pi(u,w) = \iint\delta(u-y+X)\sum_{i=1}^N\delta(w-X+x_i)P\Theta(y-X)\,dx^Ndy.\label{eq:Pi-def}
\end{equation}
Both correlations functions $\rho^{(2)}(s,s')$ and $\Pi(u,w)$ quantify nontrivial correlations in the density  of filament tips in the gel, but differently. The first of these has the form of a familiar two-particle density function, and is symmetric in its arguments. The second quantifies correlations between the lead filament and a general lagging filament. The two functions decay very differently for large values of their arguments: it is much more likely to find \emph{any} two filaments at a large distance from the obstacle, than to find two filaments at a large distance from the obstacle \emph{and} one  of these is the lead filament. In the latter case \emph{all} filaments are a large distance from the obstacle. For this reason, $\Pi$ generally decays much more rapidly that $\rho^{(2)}$ when both arguments are large.

The various one- and two-filament densities are connected by certain exact relationships. First, from their definitions it is easy to check that
\begin{equation}
\int_0^\infty \Pi(u,s-u)\,du = \int_0^s\Pi(u,s-u)\,du=\rho(s).\label{eq:Pi-relation-1}
\end{equation}
Second,
\begin{equation}
\Pi(0,s) =\delta(s)\rho(0) +  \rho^{(2)}(s,0).\label{eq:Pi-relation-2}
\end{equation}
We derive Eq. \eqref{eq:Pi-relation-2} from definition \eqref{eq:Pi-def} by partitioning  the domain of integration into the sets $\{X=x_k\}_{k=1}^N$ as in our analysis of $\sigma$,
\begin{equation*}
\Pi(0,s) = \sum_{k=1}^N\iint\delta(x_k-y)\sum_{i=1}^N\delta(s-x_k+x_i)P\Theta(y-x_k)\chi_{\{X=x_k\}}\,dx^Ndy.
\end{equation*}
 We find
\begin{equation}
\Pi(0,s) =\sum_{k=1}^N\iint\delta(x_k-y)\delta(s)P\Theta(y-x_k)\,dx^Ndy
+\sum_{k=1}^N\iint\delta(x_k-y)\sum_{i\ne k}\delta(s-x_k+x_i)P\Theta(y-x_k)\,dx^Ndy.\label{exp:Pi-relation-2-2}
\end{equation}
Identity \eqref{eq:Pi-relation-2} can now be read off from expression \eqref{exp:Pi-relation-2-2} and the definition of the various density functions.

\section{Dynamical equations for the densities\label{sec:exact-equations}}
\subsection{The instantaneous, ensemble averaged drift $v$}

The time evolution of the probability density $P$ obeys the master equation:
\begin{equation}
\frac{\partial P}{\partial t}=D\frac{\partial^2 P}{\partial y^2} + \sum_{i=1}^N \Theta(y-x_i+1)P(t,y,\dots,x_i -1,\dots) - \Theta(y-x_i-1)P.\label{eq:master-appendix}
\end{equation}
Eq. \eqref{eq:master-appendix} is complemented by the reflecting boundary condition 
\begin{equation}
\frac{\partial P}{\partial y}\left(\{y=x_i\}\right)=0,\quad i=1,\dots,N.\label{eq:master-bc}
\end{equation}
Using  master equation \eqref{eq:master-appendix} and its boundary condition, we find an expression for the instantaneous ensemble averaged drift  velocity $v$,
\begin{equation}
v=\iint y\Theta(y-X) \frac{\partial P}{\partial t}\,dx^N dy.\label{eq:V-integral}
\end{equation}

 On substituting Eq. \eqref{eq:master-appendix} into Eq. \eqref{eq:V-integral}, the polymerization terms [the second and third terms on the right hand side of  Eq. \eqref{eq:master-appendix}]  cancel. To see this, make the change of variable  $x_i\to x'_i=x_i-1$ in the inner sum of the first polymerization term for each $i$  (this change of variable preserves the measure). Using the equivalence
\begin{equation}
\Theta(y-X)=\prod_{i=1}^N\Theta(y-x_i),\label{eq:Theta-formula}
\end{equation}
we find the equivalent expression for the first polymerization term in Eq. \eqref{eq:master-appendix}:
\[
\sum_{i=1}^N\iint y\prod_{j\ne i}\Theta(y-x_j) \Theta(y-x'_i)\Theta(y-x'_i-1)P\,dx^Ndy.
\]
 It is easy to see that this transformed integral is canceled by the second polymerization  term.

Thus the only non-vanishing contribution to the integral for $V$, Eq. \eqref{eq:V-integral}, comes from the first term on the  right hand side of Eq. \eqref{eq:master-appendix}, i.e. the diffusion of the obstacle and its obstruction by the gel.
From the first line on the right hand side of Eq. \eqref{eq:master-appendix} we compute
\begin{align}
D\iint y\Theta(y-X) \frac{\partial^2 P}{\partial y^2}\,dx^N dy&= D\iint-\frac{\partial}{\partial y}\left[y\Theta(y-X)\right]\frac{\partial P}{\partial y}\,dx^N dy \notag\\
&=D\iint-\Theta(y-X)\frac{\partial P}{\partial y}\,dx^N dy \notag\\
&=D\iint\frac{\partial }{\partial y}\left[\Theta(y-X)\right]P\,dx^N dy.\label{eq:diffusion-terms}
\end{align}
We have used the product rule and the boundary condition Eq. \eqref{eq:master-bc} in the third line. 
The final line of  expression \eqref{eq:diffusion-terms} can be put into a more useful form by recalling the definitions of  $\rho$ [Eq. \eqref{eq:rho-def}] and $\psi$ [Eq. \eqref{eq:psi-def}] and using identity \eqref{eq:Theta-formula}. We find
\[
v=D\psi(0)=D\rho(0),
\]
which is Eq. \eqref{eq:velocity-formulas} of the main text.

\subsection{Derivation of the equation for $\rho$}

Here we explain the derivation of the equations governing the time evolution of the density $\rho$. The final equation is
\begin{equation}
\frac{\partial \rho}{\partial t} = D\frac{\partial^2 \rho}{\partial s^2} + D\frac{\partial }{\partial s}\rho^{(2)}(s,0) +\rho(s+1) - \Theta(s-1)\rho(s)\label{eq:rho-eq-appendix}
 \end{equation}

\subsubsection{Diffusion and drift}
The first two terms on the right hand side of Eq. \eqref{eq:rho-eq-appendix} as well as the boundary condition for this equation can be derived by analyzing the diffusion term in the master equation [the first term on the right side of Eq. \eqref{eq:master-appendix}].

We start our derivation by expressing the time evolution of the filament tip density,
\begin{equation*}
\frac{\partial \rho}{\partial t} = \iint\sum_{j=1}^N \delta(s-y+x_j)\Theta(y-X)\frac{\partial P}{\partial t}\, dx^Ndy, 
\end{equation*}
where $\partial P/\partial t$ is to be replace by  the master equation. The contribution to $\partial \rho/\partial t$ from the diffusion term in Eq. \eqref{eq:master-appendix} is
\[
D\iint\sum_{j=1}^N \delta(s-y+x_j)\Theta(y-X)\frac{\partial^2 P}{\partial y^2}\, dx^Ndy.
\]
Picking one term in the  sum and integrating by parts once gives 
\[
-D\iint\sum_{j=1}^N \frac{\partial }{\partial  y}\left[\delta(s-y+x_j)\Theta(y-X)\right]\frac{\partial P}{\partial y}\, dx^Ndy.
\]
Using the product rule, Eq. \eqref{eq:Theta-formula}, and a well-known property of convolution, this last expression is equivalent to
\begin{multline}
D\frac{\partial }{\partial s}\left[\iint\sum_{j=1}^N \delta(s-y+x_j)\Theta(y-X)\frac{\partial P}{\partial y}\, dx^Ndy\right] +\\
 D\iint\sum_{j=1}^N \delta(s-y+x_j)\sum_{k\ne j}\delta(y-x_k)\prod_{l\ne k}\Theta(y-x_l)\frac{\partial P}{\partial y}\, dx^Ndy+ \\
   D\iint\sum_{j=1}^N \delta(s-y+x_j)\delta(y-x_j)\prod_{l\ne j}\Theta(y-x_l)\frac{\partial P}{\partial y}\, dx^Ndy.\label{exp:diffusion-1}
\end{multline}
The second and third lines in this expression follow from the application of the derivative operator to $\Theta(y-X).$ The second line contains the term when the derivative acts on the factor in the product of Heaviside functions depending on the variable $x_j$ that also appears in the factor $\delta(s-y-x_j)$. The third line contains  the terms when the derivative acts on other factors in the product of Heaviside functions. In this way we see that the second line in  expression \eqref{exp:diffusion-1} vanishes because of the boundary condition, Eq. \eqref{eq:master-bc}. The final line of expression \eqref{exp:diffusion-1} will contribute to the boundary condition for Eq. \eqref{eq:rho-eq-appendix}.

Now integrating by parts again and using the properties of convolution a second time in the first line of expression \eqref{exp:diffusion-1}, we find for that line
\begin{multline}
D\frac{\partial^2 }{\partial s^2}\left[\iint\sum_{j=1}^N \delta(s-y+x_j)\Theta(y-X)P\, dx^Ndy\right]\\
-D\frac{\partial }{\partial s}\left[\iint\sum_{j=1}^N \delta(s-y+x_j)\sum_{k\ne j} \delta(y-x_k)\prod_{l\ne k}\Theta(y-x_l)P_{N}\, dx^Ndy\right]\\
-D\frac{\partial }{\partial s}\left[\iint\sum_{j=1}^N \delta(s-y+x_j) \delta(y-x_j)\prod_{l\ne j}\Theta(y-x_l)P_{N}\, dx^Ndy\right].\label{exp:diffusion-2}
\end{multline}
Here again we have separated the term which contributes to the boundary condition for Eq. \eqref{eq:rho-eq-appendix}.
Combining  expression \eqref{exp:diffusion-2} with the last term in expression \eqref{exp:diffusion-1}  and introducing the one- and two-filament density functions [see Eqs. \eqref{eq:rho-def} and \eqref{eq:rho2-def}] we have
\begin{multline}
D\frac{\partial^2 \rho}{\partial s^2}(s)-D\frac{\partial \rho^{(2)}}{\partial s}(s,0)
-D\frac{\partial }{\partial s}\left[\iint\sum_{j=1}^N \delta(s-y+x_j) \delta(y-x_j)\prod_{l\ne j}\Theta(y-x_l)P\, dx^Ndy\right]\\
+D\iint\sum_{j=1}^N \delta(s-y+x_j)\delta(y-x_j)\prod_{l\ne j}\Theta(y-x_l)\frac{\partial P}{\partial y}\, dx^Ndy.\label{exp:diffusion-3}
\end{multline}

We can now recognize the first two terms in expression \eqref{exp:diffusion-3} as the first two terms on the right side of Eq. \eqref{eq:rho-eq-appendix}. We will return to the final two terms in expression \eqref{exp:diffusion-3} when we derive the boundary condition for Eq. \eqref{eq:rho-eq-appendix}

\subsubsection{Polymerization}
We next turn to the contribution to the equation for $\rho(s)$ from the polymerization terms in the master equation, Eq. \eqref{eq:master-appendix}. To compute these, we must evaluate the integral
\begin{equation}
\iint \left[\sum_{j=1}^N \delta(s-y+x_j)\right]\left[\sum_{i=1}^N \Theta(y-x_i+1)P(\dots,x_i-1,\dots) - \Theta(y-x_i-1)P\right]dx^Ndy.
\end{equation}
We expand the product of sums and separate the terms for which $j=i$ from the terms for which $j\ne i$. In the latter case, we obtain expressions of the form
\begin{equation}
\iint \delta(s-y+x_j)\left[ \Theta(y-x_i+1)P(\dots,x_i-1\dots) - \Theta(y-x_i-1)P\right]dx^Ndy.
\end{equation}
After making the change of variable $x_i^\prime = x_i - 1$ in the first term in brackets, it is easy to see that such terms integrate to zero. This leaves the terms with $i=j$, these are:
\begin{equation}
\iint \sum_{j=1}^N \delta(s-y+x_j)\left[\Theta(y-x_j+1)P(\dots,x_j-1,\dots) - \Theta(y-x_j-1)P\right]dx^Ndy.
\end{equation}
Making the change of variable $x'_j = x_j - 1$, this integral can be written as
\begin{equation}
\iint \sum_{j=1}^N \delta(s-y+x'_j+1)\Theta(y-x'_j)P - \delta(s-y+x_j)\Theta(y-x_j-1)Pdx^Ndy.
\end{equation}
Then using the $\delta$-functions, we rewrite the $\Theta$-functions in terms of the variable $s$ and compute
\begin{equation}
\iint \sum_{j=1}^N \delta(s-y+x'_j+1)\Theta(s+1)P - \delta(s-y+x_j)\Theta(s-1)Pdx^Ndy.
\end{equation}
We can express this integral in  terms of densities as
\begin{equation}
\Theta(s+1)\rho(s+1) - \Theta(s-1)\rho(s).\label{eq:rho-eq-polymerization-part}
\end{equation}
(Of course $s>1$ so the first $\Theta$-function is unity.)

\subsubsection{Boundary conditions for $\rho$}

We now add expressions \eqref{exp:diffusion-3} and \eqref{eq:rho-eq-polymerization-part} to complete the derivation of Eq. \eqref{eq:rho-eq-appendix} and its boundary condition. We begin by selecting any test function $\zeta(s)$ compactly supported in $(0,\infty)$, i.e. which vanishes on some open interval of $s=0$. We multiply both expressions \eqref{exp:diffusion-3} and \eqref{eq:rho-eq-polymerization-part} by $\zeta$, integrate with respect to $s$, and add the results. Our choice of test function removes the boundary terms in expression \eqref{exp:diffusion-3}, and we find
\[
\int_0^\infty\zeta(s)\left\{\frac{\partial \rho}{\partial t} -\left[ D\frac{\partial^2 \rho}{\partial s^2} + D\frac{\partial }{\partial s}\rho^{(2)}(s,0) +\rho(s+1) - \Theta(s-1)\rho(s)\right]\right\}ds=0.
\]
As this equation holds for any test function which is supported on $(0,\infty)$, we conclude that Eq. \eqref{eq:rho-eq-appendix} holds almost everywhere.

To derive the boundary conditions, we choose a different test function $\eta(s)$ which is nonzero at $s=0$, but who's first derivative $d\eta/ds$ is compactly supported in $(0,\infty)$ (i.e. it vanishes in a neighborhood of $s=0$). Using this test function and Eq. \eqref{eq:rho-eq-appendix} we find that the boundary terms in expression \eqref{exp:diffusion-3} contribute
\begin{multline}
\int_0^\infty \eta(s)\frac{\partial }{\partial s}\left[\iint\sum_{j=1}^N \delta(s-y+x_j) \delta(y-x_j)\prod_{l\ne j}\Theta(y-x_l)P\, dx^Ndy\right]ds=\\
\int_0^\infty\eta(s)\iint\sum_{j=1}^N \delta(s-y+x_j)\delta(y-x_j)\prod_{l\ne j}\Theta(y-x_l)\frac{\partial P}{\partial y}\, dx^Ndyds.\label{eq:rho-u-bc-1}
\end{multline}
Notice that we can transform the right side of Eq. 
\eqref{eq:rho-u-bc-1} to  
\begin{multline}
\int_0^\infty \eta(s)\frac{\partial }{\partial s}\left[\iint\sum_{j=1}^N \delta(s-y+x_j)\delta(y-x_j)\prod_{l\ne j}\Theta(y-x_l)P\, dx^Ndy\right]ds=\\
\int_0^\infty\eta(s)\delta(s)\iint\sum_{j=1}^N \delta(s-y+x_j)\prod_{l=1}^N\Theta(y-x_l)\frac{\partial P}{\partial y}\, dx^Ndyds.\label{eq:rho-bc-2}
\end{multline}
Here, we have used the first $\delta$-function (which depends on $s$) to simplify the second $\delta$-function, and we have introduced a new $\Theta$-function in the product of $\Theta$-functions to preserve the correct domain of integration.

Integrating by parts on the right hand side of Eq. \eqref{eq:rho-bc-2} gives
\begin{multline}
\int_0^\infty \eta(s)\frac{\partial }{\partial s}\left[\iint\sum_{j=1}^N \delta(s-y+x_j)\delta(y-x_j)\prod_{l\ne j}\Theta(y-x_l)P\, dx^Ndy\right]ds=\\
\int_0^\infty\eta(s)\delta(s)\frac{\partial }{\partial s}\left[\iint\sum_{j=1}^N \delta(s-y+x_j)\prod_{l=1}^n\Theta(y-x_l)P\right]\, dx^Ndyds\\
 -\int_0^\infty\eta(s)\delta(s)\iint\sum_{j=1}^N \delta(s-y+x_j)\left[\sum_{k=1}^N\delta(y-x_k)\prod_{l\ne k}\Theta(y-x_l)\right]P\, dx^Ndyds.\label{eq:rho-bc-3}
\end{multline}
Integrating by parts on the left hand side of Eq. \eqref{eq:rho-bc-3}  and using the condition $\frac{d \eta}{ds}(0)=0$ we find
\begin{multline}
- \eta(0)\left[\iint\sum_{j=1}^N \delta(y+x_j)\delta(y-x_j)\prod_{l\ne j}\Theta(y-x_l)P\, dx^Ndy\right]=\\
\eta(0)\frac{\partial }{\partial s}\left[\iint\sum_{j=1}^N \delta(-y+x_j)\prod_{l=1}^N\Theta(y-x_l)P\right]\, dx^Ndy\\
 -\eta(0)\iint\sum_{j=1}^N \delta(-y+x_j)\left[\sum_{k=1}^N\delta(y-x_k)\prod_{l\ne k}\Theta(y-x_l)\right]P\, dx^Ndy.\label{eq:rho-bc-4}
\end{multline}
The term on the left of Eq. \eqref{eq:rho-bc-4} cancels a term on the right leaving exactly
\begin{multline}
0=\frac{\partial }{\partial s}\left[\iint\sum_{j=1}^N \delta(-y+x_j)\prod_{l=1}^N\Theta(y-x_l)P\right]\, dx^Ndy\\
 -\iint\sum_{j=1}^N \delta(-y+x_j)\left[\sum_{k=\ne j}\delta(y-x_k)\prod_{l\ne k}\Theta(y-x_l)\right]P\, dx^Ndy.
\end{multline}
This equation can be written compactly in terms of densities as
\begin{equation}
\frac{\partial \rho}{\partial s}(0)=\rho^{(2)}(0,0),
\end{equation}
which is the boundary condition for Eq. \eqref{eq:rho-eq-appendix}.

\subsection{Derivation of the equation for $\psi$}
The time evolution of the distribution $\psi$ [see Eq. \eqref{eq:psi-def}], which characterizes the location of the lead filament, can be computed as 
\begin{equation}
\frac{\partial \psi}{\partial t} = \iint \frac{\partial P}{\partial t}\delta(u-y+X)\Theta(y-X)dx^N dy.\label{eq:psi-eq-1}
\end{equation}
We now substitute the master equation for $\partial P/\partial t$ and write the result in terms of the densities $\psi$ and $\Pi$ [see Eq. \eqref{eq:Pi-def}].

\subsubsection{Diffusion}
We first consider the contribution to $\partial \psi/\partial t$ coming from the  the diffusion terms in Eq. \eqref{eq:master-appendix}. This contribution is easily evaluated by integrating by parts twice. The calculation parallels that for $\rho$, which was presented in greater detail. On the first  integration we find
\begin{multline}
D\iint \frac{\partial^2 P}{\partial y^2}\delta(u-y+X)\Theta(y-X)dx^N dy=\\
D\frac{\partial }{\partial u}\iint \frac{\partial P}{\partial y}\delta(u-y+X)\Theta(y-X)dx^N dy 
- D\iint \frac{\partial P}{\partial y}\delta(u-y+X)\frac{\partial}{\partial y} \Theta(y-X)dx^N dy,
\end{multline}
where the second term evaluates to zero because of the boundary condition on $P$. On a second integration we find
\begin{multline}
D\iint \frac{\partial^2 P}{\partial y^2}\delta(u-y+X)\Theta(y-X)dx^N dy=\\
D\frac{\partial^2 }{\partial u^2}\iint  P\delta(u-y+X)\Theta(y-X)dx^N dy
 - \delta(u)D\frac{\partial }{\partial u}\iint P\delta(u-y+X) \Theta(y-X)dx^N dy.
\end{multline}
Writing this in terms of the  distribution $\psi$, we find
\begin{equation}
D\frac{\partial^2 \psi}{\partial u^2} + \delta(u)D\frac{\partial \psi}{\partial u},\label{exp:psi-diffusion}
\end{equation}
in which the second term furnishes the boundary condition on $\psi$, i.e. $\frac{\partial \psi}{\partial u}(0)=0$ .

\subsubsection{Polymerization}
To compute the contribution to $\partial \psi/\partial t$ from polymerization, we must evaluate the integral
\begin{equation}
\iint \delta(u-y+X)\left[\sum_{i=1}^N \Theta(y-x_i+1)P(\dots,x_i-1,\dots)- \Theta(y-x_i-1)P\right]\,dx^Ndy\label{exp:psi-polymerization-1}
\end{equation}
To evaluate this expression, for each $i$, we separate the domain of integration in the first term of expression \eqref{exp:psi-polymerization-1} into the sets $\{x_i=X\}$ and $\{x_i<X\}$.  Considering the second case when  $x_i<X$, we introduce a factor of $\Theta(X-x_i)$ for each $i$, and we have 
\begin{equation}
\iint \delta(u-y+X)\Theta(y-x_i+1)P(\dots,x_i-1,\dots)\Theta(X-x_i)\,dx^Ndy.\label{exp:psi-polymerization-2}
\end{equation}
We define the new variable $x'_i = x_i - 1$, and on noticing that $X=\max \{x_{j\ne i},x'_i\}$ (because $x_i$ is not the lead filament), we see that  we can immediately redefine $x'_i\to x_i$ so that expression \eqref{exp:psi-polymerization-2} becomes
\begin{equation}
\iint \delta(u-y+X)\Theta(y-x_i)\Theta(X-x_i-1)P\,dx^Ndy.\label{exp:psi-polymerization-3}
\end{equation}
This expression is partially canceled by the second term in brackets in expression \eqref{exp:psi-polymerization-1}, which is
\begin{equation}
- \iint \delta(u-y+X)\Theta(y-x_i-1)P\,dx^Ndy.\label{exp:psi-polymerization-4}
\end{equation}
Notice that since $y>X>x_i+1$ in expression \eqref{exp:psi-polymerization-3}, we can make the replacement $\Theta(y-x_i)\to \Theta(y-x_i-1)$ in expression \eqref{exp:psi-polymerization-3} without changing the value of this expression.  Then adding the result to  expression \eqref{exp:psi-polymerization-4} we find
\begin{equation}
-\iint \delta(u-y+X)\Theta(y-x_i-1)\Theta(x_i+1 - X)P\,dx^Ndy.\label{exp:psi-polymerization-5}
\end{equation}
(Notice the sign of the expression and the argument of the second $\Theta$-function).

We convert  expression \eqref{exp:psi-polymerization-5} into an expression involving  $\psi$ and $\Pi$ by introducing a $\delta$-function, the dummy variable $w$, and integrating with respect to $w$,
\begin{multline}
-\int_0^\infty\iint \delta(u-y+X)\delta(w-X+x_i)\Theta(y-x_i-1)\Theta(x_i+1 - X)P\,dx^Ndydw=\\
-\int_0^\infty\Theta(u+w-1)\Theta(1-w)\iint \delta(u-y+X)\delta(w-X+x_i)P\,dx^Ndydw
 \label{exp:psi-polymerization-6}
\end{multline}
By summing on $i$ and writing the result in terms of $\Pi$ [see definition \eqref{eq:Pi-def}] we have
\begin{equation}
-\int_0^\infty \Theta(u+w-1)\Theta(1-w)\Pi(u,w)\,dw=-\int_{1-u}^1\Pi(u,w)dw.
\label{exp:psi-polymerization-5b}
\end{equation}

It remains to compute the contribution from the first term in expression \eqref{exp:psi-polymerization-1} when $x_i=X$. In this case we introduce the characteristic function $\chi_{\{x_i=X\}}$ to constrain the domain of integration to the sets when the $i^{th}$ filament is the lead filament, $\{x_i=X\}$,
\begin{equation}
\iint \delta(u-y+X)\Theta(y-x_i+1)P(\dots,x_i-1,\dots)\chi_{\{X=x_i\}}\,dx^Ndy.\label{exp:psi-polymerization-6}
\end{equation}
We next define $x'_i=x_i-1$ and $X_i\equiv\max\{x_{j\ne i},x'_i\}$ and notice that $X=x'_i +1$ because of the characteristic function. Furthermore, we can write 
\[
\chi_{\{X=x_i\}} =\Theta(x'_i+1-X_i). 
\]
Then using this new notation, expression \eqref{exp:psi-polymerization-6} becomes  
\begin{equation}
\iint \delta(u-y - X_i + X_i+x'_i +1)\Theta(y-x'_i)P\Theta(x'_i + 1 -X_i)\,dx^Ndy.\label{exp:psi-polymerization-7}
\end{equation}
We convert this to an expression involving densities by introducing a $\delta$-function and dummy variable $w$ as
\begin{multline}
\int_0^\infty\iint \delta(u-y + X_i - X_i+x'_i +1)\Theta(y-x'_i)\delta(w-X_i+x'_i)P\Theta(x'_i + 1 -X_i)\,dx^Ndydw=\\
\int_0^\infty\Theta(u+1)\Theta(1-w)\iint \delta(u-y + X_i -w +1)\delta(w-X_i+x'_i)P\,dx^Ndydw.\label{exp:psi-polymerization-8}
\end{multline}
By redefining $x_i^\prime \to x_i$, we observe that the definition of $X_i$ in these redefined variables  coincides with that of $X$.  Then recalling the definition of $\Pi$ [Eq. \eqref{eq:Pi-def}] and summing over $i$ expression \eqref{exp:psi-polymerization-8} becomes
\begin{equation}
\int_0^\infty\Theta(u+1)\Theta(1-w)\Pi(u-w+1,w)\,dw.\label{exp:psi-polymerization-9}
\end{equation}
[Note that $\Theta(u+1)=1$ for all $u>0$, so we can drop this factor in expression \eqref{exp:psi-polymerization-9}.]

Finally assembling the partial results in expression \eqref{exp:psi-polymerization-5b} and \eqref{exp:psi-polymerization-9} with expression \eqref{exp:psi-diffusion} we have computed
\begin{equation}
\frac{\partial \psi}{\partial t} = D\frac{\partial^2 \psi}{\partial u^2} + \int_0^1\Pi(u-w+1,w)\,dw - \int_{1-u}^1\Pi(u,w)\,dw
\end{equation}
with the boundary condition
\begin{equation}
\frac{\partial \psi}{\partial u}(0) = 0.
\end{equation}


\end{document}